\documentclass[prb,twocolumn,groupedaddress,shownopacs,amssymb]{revtex4-2}
\usepackage{graphicx,psfrag,amsmath,amssymb}

\usepackage[usenames, dvipsnames]{color}

\usepackage{hyperref} 
\hypersetup{
    colorlinks=true,
    linkcolor=Blue,
    citecolor=Blue,
    filecolor=Blue,      
    urlcolor=Blue,
    pdftitle={Disordered BCS-BEC},
    }

\begin{document}

\title{Quenched disorder and the BCS-BEC crossover in the Hubbard model}

\author{M. Iskin}
\affiliation{
Department of Physics, Ko\c{c} University, Rumelifeneri Yolu, 
34450 Sar\i yer, Istanbul, T\"urkiye
}

\date{\today}

\begin{abstract}

We study the impact of weak quenched disorder on the BCS-BEC crossover in the 
Hubbard model within a functional-integral approach. By deriving the 
thermodynamic potential up to second order in both the disorder potential 
and pairing fluctuations, we obtain self-consistent expressions for the 
number equation, condensate fraction, superfluid fraction and sound speed 
at zero temperature. In the dilute BEC limit, our results analytically 
reproduce the known continuum limits of weakly-interacting bosons, where 
weak disorder depletes the superfluid more strongly than the condensate due to 
broken translational symmetry, and enhances the sound speed through the 
overcompensation of the static compressibility. These findings establish a 
unified and controlled approach for describing the BCS-BEC crossover in 
disordered lattice models, and they provide a foundation for future 
extensions to finite temperatures and multiband Hubbard models.

\end{abstract}
\maketitle

\section{Introduction}
\label{sec:intro}

The interplay of disorder and interactions represents one of the most 
fundamental and complex frontiers in condensed-matter 
physics~\cite{lee85, belitz94, saito16}. 
A central paradigm for exploring this interplay is the BCS-BEC crossover, 
i.e., the continuous evolution from a weak-coupling BCS superfluid of 
Cooper pairs to a strong-coupling BEC of tightly-bound Cooper 
molecules~\cite{ketterle08, giorgini08, strinati18, feng25}. 
While the crossover is well established in clean systems, its behavior in the 
presence of quenched disorder remains under 
scrutiny~\cite{sanchez10, nagler20, koch24}. Disorder is a 
ubiquitous feature in real materials and a controllable parameter in ultracold 
atomic gases, making its impact on condensate formation and superfluidity a 
problem of critical importance for disordered superconductors, 
strongly-correlated materials, and ultracold optical 
lattices~\cite{scalettar99, ghosal01, ospelkaus06, modugno10, 
deissler10, kondov11, jendrzejewski12, krinner13, yan17}.  

Early theoretical studies of weakly-interacting Bose gases established that 
weak disorder depletes the condensate and generates a normal component, 
leading to the suppression of superfluidity even at absolute 
zero~\cite{huang92}. 
A key result from this work is that disorder depletes the superfluid fraction 
more strongly than the condensate fraction, a finding later supported by 
analytical results for the sound velocity and phonon 
damping~\cite{giorgini94, lopatin02, cappellaro19}.
Extending beyond the bosonic limit, recent continuum-based studies 
across the BCS-BEC crossover have shown that condensate and superfluid 
depletions vary nonmonotonically with interaction strength, reaching a 
minimum near unitarity~\cite{orso07}. Anderson's theorem breaks down away 
from the BCS regime: as the system approaches the BEC regime, the superfluid 
order parameter becomes increasingly sensitive to disorder. 
Further work has also revealed microscopic disorder effects in the normal 
state above the critical temperature $T_c$, emphasizing that disorder 
strongly suppresses $T_c$ toward the BEC side by destroying phase coherence 
without breaking pairs~\cite{han11, palestini13}.

Despite these advances, a systematic and self-consistent lattice-based 
approach that captures disorder effects across the BCS-BEC crossover 
has been lacking. In this work, we address this problem by developing a 
functional-integral formalism for the disordered Hubbard model. 
We derive the thermodynamic potential up to second order in both the 
disorder potential and pairing fluctuations, providing a controlled analytic 
approach from which we obtain self-consistent expressions for the number 
equation, condensate fraction, superfluid fraction and sound speed at zero 
temperature. Importantly, in the dilute BEC limit our results reproduce the 
well-known continuum expressions for weakly-interacting bosons, validating 
the consistency of the approach. We further verify that weak disorder, by breaking 
translational symmetry, depletes the superfluid fraction more strongly than 
the condensate fraction, and enhances the sound speed through the 
overcompensation of the static compressibility.  

This work establishes a unified theoretical foundation for the BCS-BEC 
crossover in disordered lattice models and provides a basis for future 
extensions to finite temperatures and multiband Hubbard models. 
The remainder of the paper is organized as follows. In Sec.~\ref{sec:fia}, 
we develop the functional-integral treatment of the 
disordered Hubbard model and derive the corresponding thermodynamic potential. 
Section~\ref{sec:bbc} discusses the number equation and its evolution across 
the BCS-BEC crossover. In Secs.~\ref{sec:cf} and~\ref{sec:ccs}, we analyze the 
condensate and superfluid fractions, along with the collective mode spectrum 
and sound speed, with particular emphasis on the dilute BEC limit. 
Section~\ref{sec:ni} presents numerical results for a square lattice. 
Finally, Sec.~\ref{sec:conc} provides a summary and outlook, followed by an 
Appendix on the derivation of the superfluid fraction.

\section{Functional-Integral Approach}
\label{sec:fia}

In quenched disorder, the time scale of the disorder potential is assumed 
to be much longer than the thermodynamic one. This implies that the disorder
potential may depend only on space, and that the disorder average must 
be performed after the thermal one~\cite{cappellaro19}. For simplicity, 
in this paper we assume a static white-noise disorder potential
$
V_\uparrow(\mathbf{r}_i, t) = V_\downarrow(\mathbf{r}_i, t) 
\equiv V(\mathbf{r}_i),
$
which is random in space, i.e., 
$
\langle V(\mathbf{r}_i) \rangle_d = 0,
$
and characterized by a delta-function correlation
$
\langle V(\mathbf{r}_i) V(\mathbf{r}_j) \rangle_d = \kappa \delta_{ij}.
$
Here, $\sigma = (\uparrow, \downarrow)$ denotes the spin index, 
$\mathbf{r}_i$ is the position of the $i$th lattice site, 
$\langle \cdots \rangle_d$ represents the statistical average over random
disorder realizations, $\kappa \ge 0$ is the disorder strength in units of 
(energy)$^2$, and $\delta_{ij}$ is the Kronecker delta. 
In lattice momentum-Matsubara frequency space we introduce 
$
q \equiv (\mathbf{q}, iq_\ell),
$ 
with bosonic Matsubara frequencies $q_\ell = 2\ell \pi/\beta$ 
for $\ell = 0,\pm 1, \pm 2,\cdots$, where $\beta = 1/T$ is the inverse temperature 
(with $k_\mathrm{B} \to 1$ the Boltzmann constant).  
Even though we are ultimately interested in $T = 0$, it is convenient to use 
the Matsubara formalism and take the limit $\beta \to \infty$ at the end.
In this representation, the disorder correlations become
\begin{align}
\langle V_{q = 0} \rangle_d &= 0,
\\
\langle V_{q'} V_q \rangle_d &= \beta \kappa \, \delta_{q',-q} \, \delta_{q_\ell 0}.
\end{align} 
The Fourier components of the disorder potential are defined as 
$
V_q = \frac{1}{\sqrt{\beta N}} \int_0^\beta d\tau \sum_{i} 
e^{-i \mathbf{q} \cdot \mathbf{r}_i + iq_\ell \tau} V(\mathbf{r}_i),
$ 
where $N$ is the number of lattice sites and $\tau = it$ is the imaginary time.  
In addition, we define 
$
V_\mathbf{q} = \frac{1}{\sqrt{N}} \sum_{i} e^{-i \mathbf{q} \cdot \mathbf{r}_i} V(\mathbf{r}_i),
$ 
which leads to the relation 
$
V_q = \sqrt{\beta} \, V_\mathbf{q},
$
with $V_{-q} = V^*_q$ since the disorder potential is real.

\subsection{Grand Partition Function}
\label{sec:pf}

In this paper, our strategy is to calculate the grand partition function 
$\mathcal{Z}$ of the system for a particular microscopic realization 
of the disorder potential, and then average all such $\mathcal{Z}$'s 
over the random disorder realizations~\cite{orso07, cappellaro19}. 
For this reason, using the imaginary-time functional path-integral 
approach~\cite{sademelo93, engelbrecht97, taylor06, diener08, orso07}, 
we start by writing
\begin{align}
\mathcal{Z} = \int \mathcal{D}[c, \tilde{c}] \, e^{-\mathcal{S}(c, \tilde{c})},
\end{align}
where $c$ and $\tilde{c}$ are Grassmann variables obeying 
anti-periodic boundary conditions, e.g., 
$c(\tau) = -c(\tau + \beta)$, 
and $\mathcal{S}(c, \tilde{c})$ is the action. We work in momentum space 
and employ the canonical transformation
$
c_{\mathbf{k} \sigma}(\tau) = \frac{1}{\sqrt{N}} \sum_{\mathbf{r}_i}
e^{-i \mathbf{k} \cdot \mathbf{r}_i} \, c_{\mathbf{r}_i \sigma}(\tau),
$
to express the action as
\begin{align}
\mathcal{S}(c, \tilde{c}) = \int_0^\beta d\tau 
\bigg[\sum_{\mathbf{k}\sigma} \tilde{c}_{\mathbf{k} \sigma}(\tau) 
\frac{\partial}{\partial \tau} c_{\mathbf{k} \sigma}(\tau) 
+ \mathcal{H}(\tau)\bigg],
\end{align}
where
$
\mathcal{H}(\tau) = \mathcal{H}_0(\tau) + \mathcal{H}_d(\tau) 
+ \mathcal{H}_U(\tau)
$
is obtained by replacing the field operators $c^\dagger$ and $c$ 
in the Hamiltonian with the Grassmann variables $\tilde{c}(\tau)$ 
and $c(\tau)$. Here,
$
\mathcal{H}_0(\tau) = \sum_{\mathbf{k} \sigma} \xi_{\mathbf{k} \sigma}
\tilde{c}_{\mathbf{k} \sigma}(\tau) c_{\mathbf{k} \sigma}(\tau)
$
is the kinetic term with
$
\xi_{\mathbf{k} \sigma} = \varepsilon_{\mathbf{k} \sigma} - \mu,
$
where we assume time-reversal symmetry
$
\varepsilon_{\mathbf{k} \uparrow} = \varepsilon_{-\mathbf{k},\downarrow} 
\equiv \varepsilon_\mathbf{k}
$
and inversion symmetry $\varepsilon_\mathbf{k} = \varepsilon_{-\mathbf{k}}$,
and where $\mu$ is the chemical potential. The disorder term is
$
\mathcal{H}_d(\tau) = \frac{1}{\sqrt{N}} \sum_{\mathbf{k} \mathbf{k'} \sigma} 
V_{\mathbf{k-k'}} \, \tilde{c}_{\mathbf{k} \sigma}(\tau) 
c_{\mathbf{k'} \sigma}(\tau),
$
which depends on $V_{\mathbf{q}}$. We consider the standard Hubbard model 
with onsite attractive interactions between $\uparrow$ and $\downarrow$ particles,
\begin{align}
\mathcal{H}_U(\tau) = 
-U\sum_\mathbf{q} \tilde{B}_\mathbf{q}(\tau) B_\mathbf{q}(\tau),
\end{align}
where $U \ge 0$ is the interaction strength,
$
\tilde{B}_\mathbf{q}(\tau) = \frac{1}{\sqrt{N}} \sum_\mathbf{k}
\tilde{c}_{\mathbf{k} +\mathbf{q}/2, \uparrow}(\tau) 
\tilde{c}_{-\mathbf{k}+\mathbf{q}/2, \downarrow}(\tau)
$
creates pairs of particles with center-of-mass momentum $\mathbf{q}$ and
$
B_\mathbf{q}(\tau) = \frac{1}{\sqrt{N}} \sum_\mathbf{k} 
c_{-\mathbf{k}+\mathbf{q}/2, \downarrow}(\tau) \,
c_{\mathbf{k} +\mathbf{q}/2, \uparrow}(\tau)
$
annihilates them.

To decouple the quartic interaction term and introduce auxiliary bosonic 
degrees of freedom $\Phi_\mathbf{q}(\tau)$, we use the Hubbard-Stratonovich
transformation,
\begin{align}
e^{U \tilde{B}_\mathbf{q} B_\mathbf{q}}
= \int \mathcal{D}[\Phi, \Phi^*] e^{
-\frac{|\Phi_\mathbf{q}|^2}{U}
+ \frac{\Phi_\mathbf{q} \tilde{B}_\mathbf{q}
+ \Phi^*_\mathbf{q} B_\mathbf{q}}{\sqrt{N}}
},
\end{align}
which holds for any $\mathbf{q}$ and $\tau$. Note that
$
\Phi_\mathbf{q}(\tau) = \Phi_\mathbf{q}(\tau + \beta)
$
is a complex field that obeys periodic boundary conditions.
It is convenient to introduce Fourier transforms in imaginary time 
for the Grassmann, disorder and bosonic variables, e.g.,
$
c_{k \sigma} = \frac{1}{\sqrt{\beta}} \int_0^\beta d\tau \,
e^{i k_n \tau} \, c_{\mathbf{k} \sigma}(\tau),
$
where
$
k \equiv (\mathbf{k}, i k_n),
$
and $k_n = (2n+1)\pi/\beta$ is the fermionic Matsubara frequency
with $n = 0, \pm 1, \pm 2,\cdots$.
Using the two-component Nambu-Gorkov spinors
$
\boldsymbol{\tilde{\psi}}_k = (\tilde{c}_{k \uparrow}, \, c_{-k, \downarrow}),
$
and
$
\boldsymbol{\psi}_k = (c_{k \uparrow}, \, \tilde{c}_{-k, \downarrow})^\mathrm{T},
$
where $\mathrm{T}$ denotes the transpose, we obtain 
$
\mathcal{S}(c, \tilde{c}, \Phi, \Phi^*) 
= \frac{1}{U}\sum_q |\Phi_q|^2 - \sum_{kk'} 
\boldsymbol{\tilde{\psi}}_k
\mathbf{G}^{-1}_{k k'}
\boldsymbol{\psi}_{k'},
$
where
\begin{align}
\mathbf{G}^{-1}_{k k'} = 
\begin{bmatrix}
(ik_n - \xi_\mathbf{k})\delta_{k k'} - \dfrac{V_{k-k'}}{\sqrt{\beta N}} 
& \dfrac{\Phi_{k-k'}}{\sqrt{\beta N}} \\[6pt]
\dfrac{\Phi^*_{k'-k}}{\sqrt{\beta N}} 
& (ik_n + \xi_\mathbf{k})\delta_{k k'} + \dfrac{V_{k-k'}}{\sqrt{\beta N}}
\end{bmatrix}
\end{align}
is the inverse Nambu-Gorkov Green’s function in the momentum-frequency 
representation. Here 
$
\xi_\mathbf{k} = \varepsilon_\mathbf{k} - \mu
$ 
denotes the shifted dispersion. We do not specify the explicit form of 
$\xi_\mathbf{k}$, and our theory remains applicable as long as 
$
\xi_\mathbf{k} = \xi_{-\mathbf{k}}.
$ 
This condition is satisfied, for example, by the continuum model where 
the kinetic energy is $\varepsilon_\mathbf{k} = |\mathbf{k}|^2/(2m)$,
and the system size $N$ is replaced by the system volume $\mathcal{V}$ 
throughout.
For the lattice model, the dispersion relation is generally expressed as
\begin{align}
\varepsilon_\mathbf{k} = - \frac{1}{N}
\sum_{ij} t_{ij} e^{i \mathbf{k} \cdot \mathbf{r}_{ij}},
\end{align}
where $t_{ij}$ represents the hopping parameter from site $j$ to site $i$, 
and
$
\mathbf{r}_{ij} = \mathbf{r}_{j} - \mathbf{r}_{i}
$
denotes the relative position vector.

\subsection{Effective Gaussian Action}
\label{sec:ea}

Upon integrating out the fermionic degrees of freedom, we obtain
$
\mathcal{Z} = \int \mathcal{D}[\Phi,\Phi^*] \, e^{-\mathcal{S}(\Phi, \Phi^*)}
$
~\cite{sademelo93, engelbrecht97, taylor06, diener08, orso07}, 
where the action is given by
\begin{align}
\mathcal{S}(\Phi, \Phi^*) = \frac{1}{U}\sum_q |\Phi_q|^2 
- \mathrm{Tr} \ln(-\beta \mathbf{G}^{-1}).
\end{align}
This is a formally exact expression, where the trace $\mathrm{Tr}$ is 
taken over both Nambu space and momentum-frequency space. 
Here we used the standard matrix identity
$
\ln \det \mathbf{A} = \mathrm{Tr} \ln \mathbf{A}.
$
To make analytical progress with the superfluid ground state, 
we identify the stationary and uniform expectation value of 
$\langle \Phi_q \rangle$ as the saddle-point order parameter 
$\Delta$ for superfluidity, and express
\begin{align}
\Phi_q = \sqrt{\beta N}\, \Delta \delta_{q 0} + \phi_q,
\end{align}
where the complex bosonic field $\phi_q$ corresponds to the 
momentum-frequency fluctuations around $\Delta$. In this separation, 
$\Delta$ plays precisely the role of the mean-field (BCS) order parameter,
$
\Delta_i = U \langle c_{i \uparrow} c_{i \downarrow} \rangle,
$
i.e.,
$
\Delta = \frac{U}{N} \sum_i \langle c_{i \uparrow} c_{i \downarrow} \rangle
= \frac{U}{N} \sum_\mathbf{k} \langle c_{\mathbf{k} \uparrow} c_{-\mathbf{k}, \downarrow} \rangle.
$
We note that, in the presence of a disorder potential, the order parameter
need not be uniform. However, the typical coherence length is assumed to 
be much larger than the spatial extent of the disorder potential throughout
the BCS-BEC crossover. Thanks to the $U(1)$ symmetry, we may also set 
$\Delta$ to be real without loss of generality.
This separation also allows us to split the inverse Green’s function as
\begin{align}
\mathbf{G}^{-1}_{k k'} = \boldsymbol{\mathcal{G}}^{-1}_k \delta_{k k'} 
- \boldsymbol{\Sigma}_{k k'},
\end{align}
where the saddle-point contribution 
$
\boldsymbol{\mathcal{G}}^{-1}_k = 
ik_n \boldsymbol{\tau}_0 - \mathbf{H}_\mathbf{k}
$ 
depends only on the mean-field Hamiltonian matrix
$
\mathbf{H}_\mathbf{k}
= \xi_\mathbf{k} \boldsymbol{\tau}_z - \Delta \boldsymbol{\tau}_x,
$
and not on the disorder potential. In this paper, $\tau_0$ is an identity 
matrix and $\boldsymbol{\tau}_i$ with $i=(x,y,z)$ are the Pauli matrices 
in particle-hole space. 
The fluctuation and disorder contributions can be written as
\begin{align}
\boldsymbol{\Sigma}_{k k'} = 
\frac{1}{\sqrt{\beta N}}
\big(V_{k-k'} \boldsymbol{\tau}_z - \phi_{k-k'}\boldsymbol{\tau}_+ 
- \phi^*_{k'-k} \boldsymbol{\tau}_-\big),
\end{align}
where
$
\boldsymbol{\tau}_\pm = (\boldsymbol{\tau}_x \pm i \boldsymbol{\tau}_y)/2.
$

Assuming that the fluctuations are small and disorder effects are weak, 
we next expand the action $\mathcal{S}(\Phi, \Phi^*)$ in powers of the 
bosonic fluctuations and disorder 
potential~\cite{sademelo93, engelbrecht97, taylor06, diener08, orso07}.
Keeping terms up to second order, we obtain
$
-\mathrm{Tr} \ln(\mathbf{1} - \boldsymbol{\mathcal{G}} \boldsymbol{\Sigma}) =
\sum_k \mathrm{tr} (\boldsymbol{\mathcal{G}}_k \boldsymbol{\Sigma}_{k k})
+ \frac{1}{2}\sum_{kq} \mathrm{tr} 
(\boldsymbol{\mathcal{G}}_k \boldsymbol{\Sigma}_{k,k+q}
\boldsymbol{\mathcal{G}}_{k+q} \boldsymbol{\Sigma}_{k+q,k})
+ \cdots,
$ 
where the trace $\mathrm{tr}$ is taken only in particle-hole space.
We note that, although third- and fourth-order terms in this expansion 
also produce second-order disorder effects coupled to first- and second-order 
bosonic fluctuations, respectively, these contributions eventually result 
in disorder corrections higher than second order for the thermodynamic 
potential~\footnote{
In contrast to the $T = 0$ approach, where disorder couples to bosonic 
fluctuations through the second-order terms discussed in this paper, 
these contributions vanish near the critical temperature $T_c$, 
rendering higher-order terms essential in that regime~\cite{iskin25e}.
}.
While, in principle, higher-order couplings between the Gaussian 
fluctuations and the disorder potential could be included, such 
corrections are beyond the scope of the present study.
It is convenient to express the matrix elements of the 
mean-field Green’s function as
\begin{align}
\mathcal{G}^{11}_k &= \frac{u_\mathbf{k}^2}{ik_n - E_\mathbf{k}}
+ \frac{v_\mathbf{k}^2}{ik_n + E_\mathbf{k}},
\\
\mathcal{G}^{12}_k &= -\frac{u_\mathbf{k} v_\mathbf{k}}{ik_n - E_\mathbf{k}}
+ \frac{u_\mathbf{k} v_\mathbf{k}}{ik_n + E_\mathbf{k}},
\end{align}
where
$
u_\mathbf{k} = \sqrt{\tfrac{1}{2} + \tfrac{\xi_\mathbf{k}}{2E_\mathbf{k}}}
$
and
$
v_\mathbf{k} = \sqrt{\tfrac{1}{2} - \tfrac{\xi_\mathbf{k}}{2E_\mathbf{k}}}
$
are the usual coherence factors, and
$
E_\mathbf{k} = \sqrt{\xi_\mathbf{k}^2 + \Delta^2}
$
is the standard BCS quasiparticle excitation spectrum.
The remaining matrix elements follow as
$
\mathcal{G}^{22}_k = -\mathcal{G}^{11}_{-k}
$
and
$
\mathcal{G}^{21}_k = \mathcal{G}^{12}_k.
$

In deriving the effective Gaussian action, 
$\mathcal{S}_\mathrm{eff}(\Phi, \Phi^*)$, we set the coefficient of 
the first-order purely bosonic fluctuations to zero, and obtain the 
saddle-point condition
$
\Delta = \frac{U}{2 \beta N} 
\sum_k (\mathcal{G}^{12}_k+\mathcal{G}^{21}_k).
$
In Sec.~\ref{sec:tp}, we also show that this condition minimizes the usual
saddle-point thermodynamic potential. Furthermore, the first-order purely 
disorder term is proportional to $V_{q=0}$, i.e., it appears like a 
Hartree shift
$
V_0 n_\mathrm{F}^0 / \sqrt{\beta N}
$
in the thermodynamic potential of Sec.~\ref{sec:tp}, and therefore 
averages out. As a result, we effectively obtain
$
\mathcal{S}_\mathrm{eff}(\Phi, \Phi^*) = \mathcal{S}_\mathrm{F}(\Delta) 
+ \mathcal{S}_\mathrm{B}(\phi, \phi^*),
$
where
$
\mathcal{S}_\mathrm{F}(\Delta)
$
denotes the saddle-point action, and
$
\mathcal{S}_\mathrm{B}(\phi, \phi^*)
$
represents the Gaussian corrections due to bosonic fluctuations, 
given by~\cite{orso07}
\begin{align}
\mathcal{S}_\mathrm{F}(\Delta) &= \beta N \frac{\Delta^2}{U} 
- \sum_k \mathrm{tr} \ln(-\beta \boldsymbol{\mathcal{G}}^{-1}_k)
\nonumber \\
&+ \frac{1}{2 \beta N} \sum_{k q} 
\mathrm{tr}\!\big(
\boldsymbol{\mathcal{G}}_k \boldsymbol{\tau}_z 
\boldsymbol{\mathcal{G}}_{k+q} \boldsymbol{\tau}_z
\big) V_{-q} V_q,
\\
\label{eqn:SB}
\mathcal{S}_\mathrm{B}(\phi, \phi^*) &= \frac{1}{2}\sum_q \big(
\boldsymbol{\eta}^\dagger_q \mathbf{M}_q \boldsymbol{\eta}_q
+ V_{-q} \mathbf{W}^\mathrm{T}_q \boldsymbol{\eta}_q
+ V_q \boldsymbol{\eta}^\dagger_q \mathbf{W}_q
\big).
\end{align}
Here, 
$
\boldsymbol{\eta}^\dagger_q = (\phi^*_{q}, \phi_{-q})
$
is the fluctuation spinor, and
$
M^{11}_q = \frac{1}{U} + \frac{1}{\beta N} 
\sum_k \mathcal{G}^{11}_{k+q} \mathcal{G}^{22}_k
$
and
$
M^{12}_q = \frac{1}{\beta N} 
\sum_k \mathcal{G}^{12}_{k+q} \mathcal{G}^{12}_k,
$
with 
$
M^{22}_q = M^{11}_{-q} = (M^{11}_q)^*
$
and
$
M^{21}_q = M^{12}_q = (M^{12}_q)^*,
$
are the matrix elements of the inverse fluctuation propagator $\mathbf{M}_q$.
Upon summation over the fermionic Matsubara frequencies at zero
temperature ($T = 0$), 
we find~\cite{engelbrecht97, taylor06, diener08}
\begin{align}
\label{eqn:M11}
M^{11}_q &= \frac{1}{U} + \frac{1}{N}\sum_\mathbf{k}
\bigg( \frac{u^2 u'^2}{iq_\ell - E - E'} 
- \frac{v^2 v'^2}{iq_\ell + E + E'} \bigg), \\
\label{eqn:M12}
M^{12}_q &= -\frac{1}{N}\sum_\mathbf{k}
\bigg( \frac{uvu'v'}{iq_\ell - E - E'} 
- \frac{uvu'v'}{iq_\ell + E + E'} \bigg),
\end{align}
where we use the abbreviations 
$u \equiv u_\mathbf{k}$, $v \equiv v_\mathbf{k}$, 
$E \equiv E_\mathbf{k}$, 
$u' \equiv u_{\mathbf{k+q}}$, $v' \equiv v_{\mathbf{k+q}}$
and $E' \equiv E_{\mathbf{k+q}}$.
The coupling between the bosonic fluctuations and the disorder potential
is governed by
$
\mathbf{W}_q = (W_q, W_{-q})^\mathrm{T},
$
where
$
W_q = \frac{1}{\beta N} \sum_k \big(
\mathcal{G}^{22}_k \mathcal{G}^{12}_{k+q} 
- \mathcal{G}^{12}_k \mathcal{G}^{11}_{k+q}\big).
$
It turns out that the inclusion of these correction terms in 
Eq.~(\ref{eqn:SB}) is both necessary and sufficient to recover 
the correct BEC physics in the presence of weak disorder~\cite{orso07}.

\subsection{Thermodynamic Potential}
\label{sec:tp}

Next, we integrate out the Gaussian bosonic degrees of freedom and 
average over the random disorder realizations. 
The resulting effective grand partition function can be expressed as 
\begin{align}
\langle \ln \mathcal{Z}_\mathrm{eff} \rangle_d = - \beta N \Omega_\mathrm{eff},
\end{align}
where $\Omega_\mathrm{eff}$ is the effective thermodynamic potential
per lattice site. We can split 
$
\Omega_\mathrm{eff} = \Omega_\mathrm{F} + \Omega_\mathrm{B},
$ 
with the saddle-point part 
$
\Omega_\mathrm{F} = \mathcal{S}_\mathrm{F}(\Delta)/(\beta N)
$ 
further decomposed as 
$
\Omega_\mathrm{F} = \Omega_\mathrm{F}^0 + \Omega_\mathrm{F}^d,
$ 
where $\Omega_\mathrm{F}^0$ is the usual saddle-point potential 
and $\Omega_\mathrm{F}^d$ denotes the disorder correction. 
At $T=0$, which is the regime of interest in this work, we find
\begin{align}
\Omega_\mathrm{F}^0 &= \frac{\Delta^2}{U} - \frac{1}{N} \sum_\mathbf{k} 
(E_\mathbf{k} - \xi_\mathbf{k}),
\\
\Omega_\mathrm{F}^d &= -\frac{\kappa}{N^2} \sum_{\mathbf{k} \mathbf{q}} 
\frac{EE' - \xi\xi' + \Delta^2}{2EE'(E+E')},
\end{align}
where we use the abbreviations $\xi \equiv \xi_\mathbf{k}$ and 
$\xi' \equiv \xi_\mathbf{k+q}$.
Note that the saddle-point condition discussed above in Sec.~\ref{sec:ea} 
is equivalent to 
$
\frac{\partial \Omega_\mathrm{F}^0}{\partial \Delta}\big|_{\mu} = 0,
$
leading to the order parameter equation
\begin{align}
\label{eqn:spc}
1 = \frac{U}{N} \sum_\mathbf{k} \frac{1}{2E_\mathbf{k}}.
\end{align}
Thus, the usual saddle-point condition is not explicitly affected by 
disorder~\cite{orso07,han11}, and in accordance with Anderson's theorem, 
this implies that $\Delta$ remains essentially unaffected by weak disorder 
in the BCS regime for all fillings.
Furthermore, in the continuum model the $\mathbf{q}$-sum of 
$\Omega_\mathrm{F}^d$ develops an ultraviolet divergence due to the 
white-noise assumption for the disorder correlations, which necessitates 
a special renormalization procedure~\cite{orso07}. By contrast, our lattice 
model avoids this difficulty because the lattice spacing provides an 
intrinsic minimal length scale. In addition, while in the continuum case 
$\Omega_\mathrm{F}^d$ is independent of $\mu$~\cite{orso07}, in the 
lattice model it acquires an explicit $\mu$-dependence and therefore must 
be included in the number equation, as discussed in Sec.~\ref{sec:bbc}.

The bosonic contribution also has two parts,
$
\Omega_\mathrm{B} = \Omega_\mathrm{B}^0 + \Omega_\mathrm{B}^d,
$ 
where
\begin{align}
\Omega_\mathrm{B}^0 = \lim_{\beta \to \infty} \frac{1}{2\beta N} 
\sum_q \ln \det (\mathbf{M}_q/\beta)
\end{align}
corresponds to the usual zero-point bosonic 
fluctuations~\cite{engelbrecht97, taylor06, diener08, orso07}, which is 
typically negligible throughout the BCS-BEC 
crossover~\cite{engelbrecht97, taylor06, fukushima07, diener08}, 
and
\begin{align}
\Omega_\mathrm{B}^d = - \frac{1}{2\beta N} 
\sum_{q} \mathbf{W}^\mathrm{T}_q \mathbf{M}^{-1}_q \mathbf{W}_q
\langle V_{-q} V_q \rangle_d
\end{align}
is the contribution induced directly by the disorder potential~\cite{orso07}. 
Note that $\Omega_\mathrm{B}^d$ is real in general for any type of 
disorder potential. For the random disorder considered in this paper, 
we obtain
\begin{align}
\label{eqn:OBd}
\Omega_\mathrm{B}^d = - \frac{\kappa}{N} 
\sum_{\mathbf{q}} \frac{W^2_{\mathbf{q}, 0}}
{M^{11}_{\mathbf{q}, 0} + M^{12}_{\mathbf{q}, 0}},
\end{align}
where the coupling between bosonic fluctuations and the disorder potential
is determined by
\begin{align}
\label{eqn:Wq}
W_q = \frac{1}{2N} \sum_\mathbf{k} 
\frac{\Delta(\xi + \xi' - i q_\ell)}{(E+E')^2-(i q_\ell)^2}
\bigg( \frac{1}{E} + \frac{1}{E'} \bigg),
\end{align}
after setting $q_\ell = 0$. It can be shown that $\Omega_\mathrm{B}^d$ is
the leading disorder correction, i.e.,
$
\Omega_\mathrm{B}^d \gg \Omega_\mathrm{F}^d,
$
in the BEC regime. 
In Eq.~(\ref{eqn:OBd}), it is important to emphasize that
$
M^{11}_{\mathbf{q}, 0} + M^{12}_{\mathbf{q}, 0}
= \frac{1}{U} - \frac{1}{N} \sum_\mathbf{k} 
\frac{EE' + \xi\xi' - \Delta^2}{2EE'(E+E')},
$
and we do not substitute $U$ using the saddle-point condition, because 
the number equation is derived for a constant $U$, as discussed next.

\section{BCS-BEC crossover}
\label{sec:bbc}

In order for our formalism to be applicable throughout the BCS-BEC 
crossover problem, i.e., to arbitrarily large values of $U$, we need 
to solve the saddle-point condition Eq.~(\ref{eqn:spc}) self-consistently 
together with the proper number equation~\cite{nsr85, sademelo93, 
engelbrecht97, taylor06, diener08, orso07}. 
In this work, we determine the filling fraction $n = \mathcal{N}/N$ 
of particles per site from
$
n = - \frac{\partial \Omega_\mathrm{eff}}{\partial \mu}\big|_{\Delta},
$
which leads to the number equation
\begin{align}
\label{eqn:n}
n = n_\mathrm{F}^0 + n_\mathrm{F}^d + n_\mathrm{B}^0 + n_\mathrm{B}^d.
\end{align}
It is important to remark that we do not include the additional 
contributions 
$
- \frac{\partial \Omega_\mathrm{eff}}{\partial \Delta}\big|_{\mu} 
\frac{\partial \Delta}{\partial \mu}
$
in the number equation, where the postfactor
$
\frac{\partial \Delta}{\partial \mu} = 
\frac{\sum_\mathbf{k}\, \xi_\mathbf{k} / E_\mathbf{k}^3}
{\sum_\mathbf{k}\, \Delta / E_\mathbf{k}^3}
$ 
follows from Eq.~(\ref{eqn:spc}). We verified that including these 
terms does not reproduce the correct BEC physics, a situation that 
also occurs in the continuum model~\cite{taylor06}, where the prefactor 
is approximately $-\mu/\Delta$. This factor becomes large in the dilute 
BEC limit, leading to a non-self-consistent theory for the observables 
of interest in this work, such as the condensate and superfluid fractions.
Note that such a contribution from $\Omega_\mathrm{F}^0$ vanishes 
automatically due to the saddle-point condition.
In Eq.~(\ref{eqn:n}), while
\begin{align}
\label{eqn:nF0}
n_\mathrm{F}^0 = 1 - \frac{1}{N} \sum_\mathbf{k} \frac{\xi_\mathbf{k}}{E_\mathbf{k}},
\end{align}
corresponds to the usual contribution from the saddle-point action 
at $T=0$, the analytic expressions for the remaining terms are 
lengthy and not particularly illuminating. In particular, the zero-point 
contribution $n_\mathrm{B}^0$ is known to be negligible for any $U$ 
throughout the BCS-BEC 
crossover~\cite{engelbrecht97, taylor06, fukushima07, diener08}, 
while $n_\mathrm{F}^d$ and $n_\mathrm{B}^d$ provide the leading disorder 
corrections in the BCS and BEC regimes, respectively. 
We emphasize that our second-order perturbative expansion 
of the disorder potential is not valid for arbitrarily large values 
of $\kappa$. It applies only in the regime where disorder corrections 
remain small, e.g., when
$
n_\mathrm{F}^0 + n_\mathrm{B}^0 \gg n_\mathrm{F}^d + n_\mathrm{B}^d.
$

For instance, let us demonstrate how $n_\mathrm{B}^d$ reproduces the 
correct BEC physics in the dilute limit, i.e., $n \ll 1$, when the 
interaction strength is much larger than the single-particle bandwidth, 
i.e., 
$
U \gg (\max \varepsilon_\mathbf{k} - \min \varepsilon_\mathbf{k}).
$
For $\kappa = 0$, setting $n = n_\mathrm{F}^0$ gives
$
\mu_0 = \frac{U}{2}(n-1)
$
and
$
\Delta_0 = \frac{U}{2}\sqrt{n(2-n)}
$
in the BEC regime~\cite{nsr85}. For $\kappa \ne 0$, the saddle-point condition yields
$
\sqrt{\mu^2 + \Delta^2} = U/2,
$
while the $n_\mathrm{F}^0$ equation gives
$
\mu = \frac{U}{2}(n_\mathrm{F}^0 - 1),
$
which implies
$
\Delta = \frac{U}{2}\sqrt{n_\mathrm{F}^0(2 - n_\mathrm{F}^0)}.
$
Thus, in the dilute BEC limit when 
$
n \approx n_\mathrm{F}^0 + n_\mathrm{B}^d \ll 1,
$ 
with $\mu \to -U/2$, $|\mu| \gg \Delta$, and 
$
n_\mathrm{F}^0 \approx 2\Delta^2/U^2,
$ 
we approximate $W_{\mathbf{q},0}$ by its $\mathbf{q=0}$ value~\cite{orso07},  
$
W_0 = \frac{\Delta}{N} \sum_\mathbf{k} \frac{\xi_\mathbf{k}}{2E_\mathbf{k}^3} 
\approx -\frac{\Delta\mu}{2\sqrt{\mu^2+\Delta^2}^3},
$
in the low-$\mathbf{q}$ regime. In addition, we 
expand~\cite{engelbrecht97, taylor06, diener08, iskin24a}, 
\begin{align}
M^{11}_{\mathbf{q},0} + M^{12}_{\mathbf{q},0} = A + \sum_{ij} C_{ij} q_i q_j + \cdots
\end{align}
at small $\mathbf{q}$, where  
$
A = \frac{1}{U} - \frac{1}{N} \sum_\mathbf{k} \frac{\xi_\mathbf{k}^2}{2E_\mathbf{k}^3}
\approx \frac{\Delta^2}{2\sqrt{\mu^2 + \Delta^2}^3}
$
is the stationary coefficient, $q_i$ and $q_j$ denote the Cartesian 
components ($x$, $y$, $z$) of the center-of-mass momentum $\mathbf{q}$, 
and  
$
C_{ij} = \frac{1}{N} \sum_\mathbf{k} 
\frac{1}{8E_\mathbf{k}^3} 
\Big( 1 - \frac{5\Delta^2 \xi_\mathbf{k}^2}{E_\mathbf{k}^4} \Big)
\frac{\partial \xi_\mathbf{k}}{\partial k_i} 
\frac{\partial \xi_\mathbf{k}}{\partial k_j} 
\approx \frac{U (m_M^{-1})_{ij}}{16\sqrt{\mu^2+\Delta^2}^3}
$
is the kinetic coefficient. In the last step, we identify  
\begin{align}
(m_M^{-1})_{ij} = \frac{2}{U N} \sum_\mathbf{k} 
\frac{\partial \xi_\mathbf{k}}{\partial k_i} 
\frac{\partial \xi_\mathbf{k}}{\partial k_j} 
\end{align}
as the inverse effective-mass tensor of the Cooper molecules~\cite{iskin25b}.
After some algebra, these approximations lead to
\begin{align}
\label{eqn:nBd}
\frac{n_\mathrm{B}^d}{2} = \frac{\kappa_M}{N} \sum_\mathbf{q} 
\frac{n_M}{\big[2\mu_M + \tfrac{1}{2}\sum_{ij} (m_M^{-1})_{ij}q_i q_j\big]^2},
\end{align}
where we identify $\kappa_M = 4\kappa$ as the effective disorder strength, 
$n_M = n_\mathrm{F}^0/2$ as the effective filling fraction, 
$\mu_M = U_M n_M$ as the effective chemical potential,
and $U_M = 2U$ as the effective repulsion between Cooper molecules
in the dilute BEC limit.
Note that this analysis is also consistent with the zero-temperature 
coherence length~\cite{iskin24c}, 
$
(\xi_0^2)_{ij} = \frac{C_{ij}}{A} = \frac{(m_M^{-1})_{ij}}{4U_M n_M}.
$
Thus, Eq.~(\ref{eqn:nBd}) reproduces the well-known disorder correction 
for weakly-interacting atomic bosons in the dilute BEC 
limit~\cite{huang92, giorgini94, lopatin02, orso07},
i.e., we find
$
n_\mathrm{B}^d = \frac{\kappa_M n_M \sqrt{m_M}^3}{2\pi \sqrt{\mu_M}}
$ 
where $(m_M^{-1})_{ij} = \delta_{ij}/m_M$. 
In this interpretation, $n_M$ accounts for the filling of condensed 
Cooper molecules, whereas $n_\mathrm{B}^d/2$ represents the filling of 
molecules that are depleted from the condensate by disorder potential. 
The factor of $4$ in $\kappa_M$ accounts for the fact that the effective 
disorder potential for a molecule is
$
V_M(\mathbf{r}_i) = V_\uparrow(\mathbf{r}_i) + V_\downarrow(\mathbf{r}_i),
$
when the molecule is extremely small, i.e.,
$
\langle V_M(\mathbf{r}_i) V_M(\mathbf{r}_j)\rangle = 
4 \langle V(\mathbf{r}_i) V(\mathbf{r}_j)\rangle.
$
Since $\mu_M \propto U$ is large, $m_M^{-1} \propto 1/U$ is small and 
$\mathbf{q}$ is restricted to the Brillouin zone, the $\mathbf{q}$-dependent 
term in the denominator can be neglected for the lattice model. 
This yields $n_\mathrm{B}^d/2 = \kappa/(4\Delta^2)$. Consequently, 
$
\Delta^2 \approx \Delta_0^2 - \kappa U^2/(4\Delta_0^2),
$
which is valid provided the correction term is small compared to the leading 
contribution.

\section{Condensate Fraction}
\label{sec:cf}

The filling $n_c$ of condensed Cooper pairs can be defined through 
the normalization factor of the pair wave function~\cite{fukushima07},
$
\Psi(\mathbf{r}_i, \mathbf{r}_j) = \langle c_{\mathbf{r}_i \uparrow}^\dagger 
c_{\mathbf{r}_j \downarrow}^\dagger \rangle.
$
This leads to the definition of the so-called condensate fraction
$
n_c = \frac{1}{N} \sum_{ij} |\Psi(\mathbf{r}_i, \mathbf{r}_j)|^2,
$
where the filling of condensed particles is $2n_c$.
Thus, in momentum space we obtain
$
n_c = \frac{1}{N} \sum_{\mathbf{k} \mathbf{k'}} \langle 
c_{\mathbf{k'} \uparrow}^\dagger c_{\mathbf{k} \downarrow}^\dagger \rangle
\langle c_{\mathbf{k} \downarrow} c_{\mathbf{k'} \uparrow} \rangle,
$
and, in terms of Green's functions, it can be reexpressed as
\begin{align}
\label{eqn:nc}
n_c = \frac{1}{\beta^2 N} \sum_{\mathbf{k} \mathbf{k}' k_n k_m} 
G^{21}_{\mathbf{k} k_n, \mathbf{k}' k_n} 
G^{12}_{\mathbf{k}' k_m, \mathbf{k} k_m}.
\end{align}
Here,
$
G^{21}_{\mathbf{k k'}}(\tau, \tau') = 
-\langle T_\tau c_{-\mathbf{k}, \downarrow}^\dagger(\tau) 
c_{\mathbf{k'} \uparrow}^\dagger(\tau') \rangle
$
is the anomalous Green's function, with $T_\tau$ being the time-ordering 
operator, and its Fourier expansion at equal times is given by
$
G^{21}_{\mathbf{k k'}}(0, 0^+) = \frac{1}{\beta} \sum_{k_n} 
e^{i k_n 0^+} G^{21}_{\mathbf{k} k_n, \mathbf{k}' k_n}.
$
In the absence of disorder when $\kappa = 0$, it is well known that 
bosonic fluctuations give a negligible contribution to $n_c$ throughout 
the BCS-BEC crossover~\cite{fukushima07}. 
For this reason, we drop them for this observable, reset
$
\boldsymbol{\Sigma}_{k k'} = \frac{V_{k-k'}}{\sqrt{\beta N}} \boldsymbol{\tau}_z,
$
and compute $\mathbf{G}_{k k'}$ up to second order in the disorder potential.
This can be achieved by noting that
\begin{align}
\mathbf{G} &= (\boldsymbol{\mathcal{G}}^{-1} - \boldsymbol{\Sigma})^{-1} 
= (\mathbf{1} - \boldsymbol{\mathcal{G}} \boldsymbol{\Sigma})^{-1}
\boldsymbol{\mathcal{G}} 
= \boldsymbol{\mathcal{G}} + \boldsymbol{\mathcal{G}}
\boldsymbol{\Sigma} \mathbf{G},
\end{align}
where the last equality follows from expanding the previous expression 
as a geometric series, i.e.,
$
(1-x)^{-1} = \sum_{n=0}^\infty x^n,
$
valid for $|x| < 1$.
Plugging $\mathbf{G}_{k'' k'}$ recursively into the expression
$
\mathbf{G}_{k k'} = \boldsymbol{\mathcal{G}}_k\delta_{k k'} 
+ \sum_{k''} \boldsymbol{\mathcal{G}}_k 
\boldsymbol{\Sigma}_{k,k''} \mathbf{G}_{k'' k'},
$
yields the Dyson series
$
\mathbf{G}_{k k'} = \boldsymbol{\mathcal{G}}_k\delta_{k k'} 
+ \boldsymbol{\mathcal{G}}_k \boldsymbol{\Sigma}_{k k'} \boldsymbol{\mathcal{G}}_{k'}
+ \sum_{k''} \boldsymbol{\mathcal{G}}_k \boldsymbol{\Sigma}_{k k''} \boldsymbol{\mathcal{G}}_{k''}
\boldsymbol{\Sigma}_{k'' k'} \boldsymbol{\mathcal{G}}_{k'}
+ \cdots.
$

After calculating Eq.~(\ref{eqn:nc}) and averaging over random disorder 
realizations, we obtain
$
n_c = n_c^{0} + n_c^d,
$
where $n_c^{0}$ is the usual saddle-point contribution~\cite{leggett}, 
and $n_c^d$ is the contribution induced directly by the disorder potential. 
Keeping terms up to second order in the disorder potential, we find
\begin{align}
\label{eqn:nc0}
n_c^0 &= \frac{1}{N}\sum_\mathbf{k} \frac{\Delta^2}{4E_\mathbf{k}^2},
\\
n_c^d &= 
\frac{\kappa}{N^2} \sum_{\mathbf{k}\mathbf{q}} 
\bigg[ \frac{\Delta^2 (2E+E')\xi(\xi+\xi')}{2E^4E'(E+E')^2}
- \frac{\Delta^2}{2E^2E'(E+E')}
\bigg]
\nonumber \\
&\quad+ \frac{\kappa}{N^2} \sum_{\mathbf{k}\mathbf{q}} 
\frac{\Delta^2 (\xi+\xi')^2}{4E^2E'^2(E+E')^2}.
\label{eqn:ncd}
\end{align}
In Eq.~(\ref{eqn:ncd}), the first line originates from the cross coupling 
between the zeroth-order terms and the second-order corrections, whereas the 
second line arises from the cross coupling of two first-order corrections. 
The latter contribution plays an important role in the BCS regime, but it is 
missing in Ref.~\cite{orso07}, where the disorder average is likely performed 
at the Green's function level before evaluating Eq.~(\ref{eqn:nc}). 
For instance, in the BCS regime of the continuum model, one can show that the 
total $\mathbf{q = 0}$ contribution to $n_c^d$ vanishes, indicating that the 
first and second lines of Eq.~(\ref{eqn:ncd}) compete with each other. 
This follows from 
$
\sum_\mathbf{k} \xi_\mathbf{k}^2/E_\mathbf{k}^6 \approx \pi D_\mathrm{F}/(8\Delta^3)
$
and 
$
\sum_\mathbf{k} 1/E_\mathbf{k}^4 \approx 3\pi D_\mathrm{F}/(8\Delta^3),
$ 
where $D_\mathrm{F}$ is the density of states at the Fermi energy. 
As a result, in contrast to the numerical results of Ref.~\cite{orso07}, 
the correct disorder correction may increase monotonically from zero 
in the BCS regime to large values in the BEC regime.

Furthermore, in the BEC regime, one finds that $n_c^d$ is much smaller 
than the intrinsic disorder correction already contained in 
$n_c^0$~\cite{orso07}. 
In this regime,
$
n_c^0 \approx \Delta^2/[4(\mu^2+\Delta^2)], 
$
which can be rewritten as 
$
n_c^0 - n_{c,0} = (\Delta^2-\Delta_0^2)/U^2, 
$
valid for any filling, yielding $-\kappa/(4\Delta_0^2)$ in the dilute limit. 
Here $n_{c,0} = n(2 - n)/4$ denotes the condensate fraction in the absence 
of disorder when $\kappa = 0$. We emphasize that $|n_c^0 - n_{c,0}| \gg |n_c^d|$ 
in this regime. Thus, we confirm that the presence of a disorder 
potential leads to an additional depletion of the condensate, 
$n_{c,0} - n_c \to n_\mathrm{B}^d / 2$, in the dilute BEC limit, 
in agreement with the number equation discussed in Sec.~\ref{sec:bbc}.

\section{Current-carrying Superfluid}
\label{sec:ccs}

To derive the superfluid fraction, we follow the continuum recipe and 
introduce a so-called phase twist to the bosonic 
field~\cite{fisher93, taylor06},
$
\Phi(x) \to \Phi(x) e^{i2\mathbf{Q} \cdot \mathbf{r}_i},
$
in the inverse Green’s function $\mathbf{G}^{-1}(x,x')$ written in 
the position-time representation, where
$
x \equiv (\mathbf{r}_i, \tau).
$
We then remove this spatial dependence from the phase of the bosonic 
field by means of a unitary transformation $\mathbf{U}(\mathbf{r}_i)$,
whose matrix elements are given by
$
U_{11}(\mathbf{r}_i) = U_{22}^*(\mathbf{r}_i) 
= e^{i \mathbf{Q} \cdot \mathbf{r}_i}
$
and 
$
U_{12}(\mathbf{r}_i) = 0 = U_{21}(\mathbf{r}_i).
$
In momentum-frequency space, this procedure is equivalent to replacing
$
\boldsymbol{\mathcal{G}}^{-1}_k \to \boldsymbol{\bar{\mathcal{G}}}^{-1}_{\bar{k}} 
= i\bar{k}_n \boldsymbol{\tau}_0 - \mathbf{\bar{H}}_\mathbf{k},
$ 
where
$
\bar{k} \equiv (\mathbf{k}, i\bar{k}_n)
$
with
$
i\bar{k}_n = ik_n - \frac{\xi_{\mathbf{k+Q}} - \xi_{\mathbf{k-Q}}}{2},
$
and
$
\mathbf{\bar{H}}_\mathbf{k}
= \bar{\xi}_\mathbf{k} \boldsymbol{\tau}_z - \Delta \boldsymbol{\tau}_x,
$
where
$
\bar{\xi}_\mathbf{k} = \frac{\xi_{\mathbf{k+Q}} + \xi_{\mathbf{k-Q}}}{2}.
$
It is convenient to express the matrix elements of the Green’s function as
\begin{align}
\bar{\mathcal{G}}^{11}_{\bar{k}} &= \frac{\bar{u}_\mathbf{k}^2}{i\bar{k}_n - \bar{E}_\mathbf{k}}
+ \frac{\bar{v}_\mathbf{k}^2}{i\bar{k}_n + \bar{E}_\mathbf{k}},
\\
\bar{\mathcal{G}}^{12}_{\bar{k}} &= -\frac{\bar{u}_\mathbf{k} \bar{v}_\mathbf{k}}
{i\bar{k}_n - \bar{E}_\mathbf{k}}
+ \frac{\bar{u}_\mathbf{k} \bar{v}_\mathbf{k}}{i\bar{k}_n + \bar{E}_\mathbf{k}},
\end{align}
where
$
\bar{E}_\mathbf{k} = \sqrt{\bar{\xi}_\mathbf{k}^2 + \Delta^2},
$
$
\bar{u}_\mathbf{k} = \sqrt{\tfrac{1}{2} 
+ \tfrac{\bar{\xi}_\mathbf{k}}{2\bar{E}_\mathbf{k}}}
$
and
$
\bar{v}_\mathbf{k} = \sqrt{\tfrac{1}{2} 
- \tfrac{\bar{\xi}_\mathbf{k}}{2\bar{E}_\mathbf{k}}}.
$
The remaining matrix elements follow as
$
\bar{\mathcal{G}}^{22}_{\bar{k}} = -\bar{\mathcal{G}}^{11}_{-\bar{k}}
$
and
$
\bar{\mathcal{G}}^{21}_{\bar{k}} = \bar{\mathcal{G}}^{12}_{\bar{k}}.
$

\subsection{Superfluid Fraction}
\label{sec:sf}

The superfluid fraction is defined as the lowest-order change in the 
free energy ($\mathcal{F} = \Omega + \mu n$) of the system due to the additional 
kinetic energy induced by the imposed superflow~\cite{scalapino92}. In our self-consistent 
approach, we expand the thermodynamic potential of the current-carrying 
superfluid state as~\cite{taylor06, orso07}
\begin{align}
\label{eqn:Oeff}
\bar{\Omega}_\mathrm{eff} = \Omega_\mathrm{eff} 
+ \frac{1}{2} \sum_{ij} D_s^{ij} Q_i Q_j + \cdots,
\end{align}
up to second order in $\mathbf{Q}$, and identify
$
D_s^{ij} = \frac{\partial^2 \bar{\Omega}_\mathrm{eff}}
{\partial Q_i \partial Q_j}\big|_{\mathbf{Q = 0}}
$
as the elements of the superfluid-fraction tensor. 
We note that the standard definition of the so-called superfluid-weight 
tensor (also called superfluid-stiffness tensor) is given per unit volume, 
whereas in this paper the superfluid-fraction tensor is defined per 
lattice site. 
Thus, to determine $D_s^{ij}$, we need to repeat the calculations 
of Sec.~\ref{sec:fia}, but substitute $\boldsymbol{\mathcal{G}}_k$ 
with $\boldsymbol{\bar{\mathcal{G}}}_{\bar{k}}$ and 
$\boldsymbol{\mathcal{G}}_{k+q}$ with 
$\boldsymbol{\bar{\mathcal{G}}}_{\bar{k}+\bar{q}}$, 
where
$
\bar{q} \equiv (\mathbf{q}, i\bar{q}_\ell),
$
with
$
i\bar{q}_\ell = iq_\ell - \frac{1}{2} \big(\xi_\mathbf{k-Q} - \xi_\mathbf{k+Q} 
- \xi_\mathbf{k+q-Q} + \xi_\mathbf{k+q+Q}\big).
$
Upon summing over the fermionic Matsubara frequencies, the resultant 
expressions for $\mathbf{\bar{M}}_q$ and $\bar{W}_q$ are obtained by 
replacing $iq_\ell$ with $i\bar{q}_\ell$, 
$\xi_\mathbf{k}$ with $\bar{\xi}_\mathbf{k}$,
$E_\mathbf{k}$ with $\bar{E}_\mathbf{k}$,
$u_\mathbf{k}$ with $\bar{u}_\mathbf{k}$ and
$v_\mathbf{k}$ with $\bar{v}_\mathbf{k}$,
in Eqs.~(\ref{eqn:M11}), (\ref{eqn:M12}), and (\ref{eqn:Wq}).
The remaining matrix elements follow as
$
\bar{M}^{22}_{\bar{q}} = \bar{M}^{11}_{-\bar{q}}
$
and
$
\bar{M}^{21}_{\bar{q}} = \bar{M}^{12}_{\bar{q}}.
$

Furthermore, up to second order in $\mathbf{Q}$, the entire procedure 
of extracting $\bar{\Omega}_\mathrm{eff}$ from $\Omega_\mathrm{eff}$ 
is equivalent to substituting $iq_\ell$ and $\mu$ with
\begin{align}
\label{eqn:iql}
i\bar{q}_\ell &= iq_\ell - \sum_i s_\mathbf{k q}^i Q_i, \\
\bar{\mu} &= \mu - \frac{1}{2}\sum_{ij} p_\mathbf{k q}^{ij} Q_i Q_j,
\end{align}
respectively, in $\mathbf{M}_q$ and $W_q$, where
$
s_\mathbf{k q}^i = \frac{\partial \xi_{\mathbf{k+q}}}{\partial k_i}
- \frac{\partial \xi_\mathbf{k}}{\partial k_i}
$,
and
$
p_\mathbf{k q}^{ij} = \frac{\partial^2 \xi_\mathbf{k}}{\partial k_i \partial k_j}
$
when $\mu$ appears in $\xi_\mathbf{k}$, while
$
p_\mathbf{k q}^{ij} = \frac{\partial^2 \xi_{\mathbf{k+q}}}{\partial k_i \partial k_j}
$
when $\mu$ appears in $\xi_{\mathbf{k+q}}$.
Within this procedure, $\xi_\mathbf{k}$ and $\xi_\mathbf{k+q}$ 
are effectively replaced by 
$
\bar{\xi}_\mathbf{k} \equiv \varepsilon_\mathbf{k} - \bar{\mu}
$ 
and
$
\bar{\xi}_\mathbf{k+q} \equiv \varepsilon_\mathbf{k+q} - \bar{\mu},
$ 
an approximation valid up to second order in $\mathbf{Q}$.
As a result, since the $\mathbf{Q}$ dependence of the thermodynamic 
potential can be expressed as
$
\bar{\Omega}_\mathrm{eff} \equiv \bar{\Omega}_\mathrm{eff}(\bar{\mu}, \mathbf{Q}),
$
with 
$
\frac{\partial \bar{\mu}}{\partial Q_i} \big|_{\mathbf{Q = 0}} = 0,
$
we decompose the superfluid fraction into two parts,
\begin{align}
D_s^{ij} = D_0^{ij} - D_n^{ij},
\end{align}
where
$
D_0^{ij} = \frac{\partial \bar{\Omega}_\mathrm{eff}} {\partial \bar{\mu}} 
\frac{\partial^2 \bar{\mu}}{\partial Q_i \partial Q_j} \big|_{\mathbf{Q = 0}}
$
represents the contribution from the total particle fraction, and
$
D_n^{ij} = - \frac{\partial^2 \bar{\Omega}_\mathrm{eff}}
{\partial Q_i \partial Q_j}\big|_{\bar{\mu}, \mathbf{Q = 0}}  
$
represents the contribution from the normal fraction. This yields
\begin{align}
D_0^{ij} = \frac{1}{N} \sum_\mathbf{k} n(\mathbf{k}) [*] p_\mathbf{k q}^{ij},
\label{eqn:D0}
\end{align}
where $n(\mathbf{k})$ is the momentum distribution at $\mathbf{Q = 0}$, 
i.e., in the absence of a superflow, since
$
\frac{\partial \bar{\Omega}_\mathrm{eff}} 
{\partial \bar{\mu}} \big|_{\mathbf{Q = 0}} 
\equiv \frac{\partial \Omega_\mathrm{eff}} {\partial \mu} \big|_\Delta.
$
Here, we define a special $[*]$ operation, which means that, 
depending on whether the $\bar{\mu}$ derivative acts on $\xi_\mathbf{k}$ 
or on $\xi_{\mathbf{k+q}}$, the corresponding term in $n(\mathbf{k})$ 
must be multiplied by the associated factor $p_\mathbf{k q}^{ij}$.
By construction, 
$
n = \frac{1}{N} \sum_\mathbf{k} n(\mathbf{k})
$
gives the total particle filling. Note that 
$
D_0^{ij} = \frac{n}{m} \delta_{ij}
$
for all $U \ne 0$~\cite{Fetter}, with 
$
n = \mathcal{N} / \mathcal{V}
$
denoting the particle density, for the continuum dispersion 
$\varepsilon_\mathbf{k} = |\mathbf{k}|^2/(2m)$, 
where 
$
p_\mathbf{k q}^{ij} = \delta_{ij}/m
$
does not depend on where $\mu$ appears.
At $T = 0$ and when $\kappa = 0$, it is well known that $D_s^{ij}$ receives 
a negligible contribution from $\bar{\Omega}_\mathrm{B}^0$ throughout the 
BCS-BEC crossover~\cite{fukushima07}. For this reason, we omit it in this observable.

When $\kappa = 0$, we calculate the usual contribution $D_{s,0}^{ij}$ 
as follows. To perform the low-$\mathbf{Q}$ expansion of the thermodynamic
potential, we split the Green's function
$
\boldsymbol{\bar{\mathcal{G}}}_{\bar{k}} = \boldsymbol{\mathcal{G}}_k 
- \boldsymbol{\bar\Sigma}_\mathbf{Q},
$
where it is sufficient to retain 
$
\boldsymbol{\bar\Sigma}_\mathbf{Q} = 
\sum_i \frac{\partial \mathbf{H}_\mathbf{k}}
{\partial k_i} Q_i \boldsymbol{\tau}_z
+ \frac{1}{2} \sum_{ij} \frac{\partial^2 
\mathbf{H}_\mathbf{k}}
{\partial k_i \partial k_j} Q_i Q_j \boldsymbol{\tau}_z
$
accurately up to second order in $\mathbf{Q}$. This leads to
$
\bar{\Omega}_\mathrm{F}^0 = \Omega_\mathrm{F}^0 
+ \frac{1}{\beta N} \sum_{n=1}^\infty \frac{(\boldsymbol{\mathcal{G}}_k 
\boldsymbol{\bar\Sigma}_\mathbf{Q})^n}{n}
$
for the thermodynamic potential, from which we extract
$
D_{s,0}^{ij} = \frac{1}{\beta N} \sum_k \mathrm{tr} 
\big(
\boldsymbol{\mathcal{G}}_k
\frac{\partial^2 \mathbf{H}_\mathbf{k}}
{\partial k_i \partial k_j}
\big)
+ \frac{1}{\beta N} \sum_k \mathrm{tr} \big(
\boldsymbol{\mathcal{G}}_k
\frac{\partial \mathbf{H}_\mathbf{k}} {\partial k_i} \boldsymbol{\tau}_z
\boldsymbol{\mathcal{G}}_k
\frac{\partial \mathbf{H}_\mathbf{k}} {\partial k_j} \boldsymbol{\tau}_z
\big)
$
as the superfluid fraction. Upon summing over the fermionic Matsubara 
frequencies, one finds that
$
D_{s,0}^{ij} = D_{0,0}^{ij} = \frac{1}{N} \sum_\mathbf{k} 
n_\mathrm{F}^0(\mathbf{k})
\frac{\partial^2 \xi_\mathbf{k}}{\partial k_i \partial k_j}
$
at $T=0$ for any $U \ne 0$~\cite{denteneer93}. This result implies that the 
system is entirely superfluid and the normal fraction vanishes, i.e.,
$D_{n,0}^{ij} = 0$, throughout the BCS-BEC crossover. Here
$
n_\mathrm{F}^0(\mathbf{k}) = 1 - \xi_\mathbf{k}/E_\mathbf{k}
$
denotes the mean-field momentum distribution when $n = n_\mathrm{F}^0$.

When $\kappa \ne 0$, the disorder potential naturally drives some of 
the particles out of the superfluid state, giving rise to a finite 
normal fraction even at 
$T = 0$~\cite{huang92, giorgini94, lopatin02, cappellaro19, orso07}. 
The non-conservation of the superfluid current can be regarded as a 
consequence of the broken translational invariance caused by the 
disorder~\cite{giorgini94}. 
To reveal this, we decompose the disorder contribution as
$
D_{n, d}^{ij} = D_{\mathrm{F}, d}^{ij} + D_{\mathrm{B}, d}^{ij},
$
according to their thermodynamic origin. Note that, since
$
\bar{\Omega}_\mathrm{F}^d \equiv \bar{\Omega}_\mathrm{F}^d(\bar{\mu}, \mathbf{Q})
$
and
$
D_{\mathrm{F}, d}^{ij} =
- \frac{\partial^2 \bar{\Omega}_\mathrm{F}^d}
{\partial Q_i \partial Q_j}\big|_{\bar{\mu}, \mathbf{Q = 0}},
$
we only need to keep track of the $\mathbf{Q}$ dependence arising from the
$\bar{q}_\ell$ terms when $q_\ell$ = 0. This leads to
\begin{align}
\label{eqn:bOFd}
\bar{\Omega}_\mathrm{F}^d &= \frac{\kappa}{N^2}
\sum_{\mathbf{k} \mathbf{q}}
\frac{\bar{E} + \bar{E}'}{2\bar{E}\bar{E}'}
\frac{\bar{E}\bar{E}' - \bar{\xi} \bar{\xi}' + \Delta^2}
{(i\bar{q}_\ell)^2 - (\bar{E}+\bar{E}')^2} \Big|_{q_\ell = 0},
\\
D_{\mathrm{F}, d}^{ij} &= \frac{\kappa}{N^2} \sum_{\mathbf{k} \mathbf{q}}
\frac{EE' - \xi\xi' + \Delta^2}{EE'(E+E')^3} \,
s_\mathbf{k q}^i s_\mathbf{k q}^j.
\label{eqn:DFd}
\end{align}
Similarly, the coupling between bosonic fluctuations and the disorder
potential gives rise to
$
D_{\mathrm{B}, d}^{ij} =
- \frac{\partial^2 \bar{\Omega}_\mathrm{B}^d}
{\partial Q_i \partial Q_j}\big|_{\bar{\mu}, \mathbf{Q = 0}}.
$
In Appendix~\ref{sec:app}, by carefully tracing the $\mathbf{Q}$ dependence 
carried by the $\bar{q}_\ell$ terms, we eventually obtain
\begin{align}
\label{eqn:bOBd}
\bar{\Omega}_\mathrm{B}^d &= - \frac{\kappa}{2 N}
\sum_{\mathbf{q}} \mathbf{\bar{W}}^\mathrm{T}_{\bar{q}}
\mathbf{\bar{M}}^{-1}_{\bar{q}} \mathbf{\bar{W}}_{\bar{q}} \Big|_{q_\ell = 0},
\\
\label{eqn:DBd}
D_{\mathrm{B}, d}^{ij} &= \frac{\kappa}{N} \sum_\mathbf{q} \bigg(
\frac{e_{2, \mathbf{q}}^{ij}}{d_{0, \mathbf{q}}}
- \frac{e_{0, \mathbf{q}} d_{2, \mathbf{q}}^{ij}}
{d_{0, \mathbf{q}}^2}
\bigg).
\end{align}
The explicit forms of $d_{0, \mathbf{q}}$, $e_{0, \mathbf{q}}$,
$d_{2, \mathbf{q}}^{ij}$, and $e_{2, \mathbf{q}}^{ij}$ are provided in the
appendix. We note that although $d_{0, \mathbf{q}}$ vanishes linearly as 
$\mathbf{q \to 0}$, both $d_{2, \mathbf{q}}^{ij}$ and $e_{2, \mathbf{q}}^{ij}$ 
vanish even faster in this limit. Consequently, 
$D_{\mathrm{F}, d}^{ij}$ and $D_{\mathrm{B}, d}^{ij}$ represent the 
leading-order disorder corrections in the BCS and BEC regimes, respectively.

\subsection{Dilute BEC limit}
\label{sec:dbl}

As an illustration, let us show how $D_{\mathrm{B}, d}^{ij}$ reproduces 
the correct BEC physics when the system is dilute, i.e., when $n \ll 1$ and 
$
U \gg (\max \varepsilon_\mathbf{k} - \min \varepsilon_\mathbf{k}).
$
In the absence of a superflow when $\mathbf{Q=0}$, and upon analytic 
continuation $iq_\ell \to \omega + i0^+$, we 
expand~\cite{engelbrecht97, taylor06, iskin24a} 
\begin{align}
M^{11}_q &= \frac{A}{2} - B\omega + \sum_{ij} \frac{C_{ij} + Q_{ij}}{2} q_i q_j 
- \frac{\mathcal{D}+R}{2}\omega^2 + \cdots,
\\
M^{12}_q &= \frac{A}{2} + \sum_{ij} \frac{C_{ij} - Q_{ij}}{2} q_i q_j 
- \frac{\mathcal{D}-R}{2}\omega^2 + \cdots,
\end{align}
up to second order in $\mathbf{q}$ and $\omega$. The coefficients $A$ 
and $C_{ij}$ are given in Sec.~\ref{sec:bbc}, while the remaining 
expansion coefficients are
$
B = \frac{1}{N} \sum_\mathbf{k} \frac{\xi_\mathbf{k}}{4E_\mathbf{k}^3},
$
$
\mathcal{D} = \frac{1}{N} \sum_\mathbf{k} \frac{\xi_\mathbf{k}^2}{8E_\mathbf{k}^5},
$
$
R = \frac{1}{N} \sum_\mathbf{k} \frac{1}{8E_\mathbf{k}^3}
$
and
$
Q_{ij} = \frac{1}{N} \sum_\mathbf{k} 
\frac{1}{8 E_\mathbf{k}^3}
\frac{\partial \xi_\mathbf{k}}{\partial k_i} 
\frac{\partial \xi_\mathbf{k}}{\partial k_j}.
$
Such a low-frequency and low-momentum expansion is valid throughout 
the BCS-BEC crossover. It can be shown that
$
\det \mathbf{M}_q \approx A \sum_{ij} C_{ij} q_i q_j - B^2 \omega^2
+ (\sum_{ij} C_{ij} q_i q_j)^2
$
reproduces the correct BEC physics, where $Q_{ij} \to C_{ij}$, 
$\mathcal{D} \to R$, and $AR \ll B^2$~\cite{engelbrecht97, taylor06, iskin24a}. 
For instance, when $\kappa = 0$, setting $\det \mathbf{M}_q = 0$ 
gives spectrum
$
\omega_\mathbf{q}^2 = \frac{A}{B^2} \sum_{ij} C_{ij} q_i q_j 
+ (\sum_{ij} \frac{C_{ij}}{B} q_i q_j)^2
$
for the collective Bogoliubov modes, where
$
A C_{ij}/B^2 = U_M n_{c,0} (m_M^{-1})_{ij},
$
with
$
n_{c,0} = n(2-n)/4,
$
and
$
C_{ij}/B = (m_M^{-1})_{ij}/2.
$
In the particular case when the system is isotropic, i.e., 
when $(m_M^{-1})_{ij} = \delta_{ij}/m_M$, we find
$
\omega_M^2 = \sqrt{\frac{U_M n_{c,0}}{m_M} |\mathbf{q}|^2 
+ \frac{|\mathbf{q}|^4}{4m_M^2}},
$
leading to
$
c_0 = \sqrt{U_M n_{c,0}/m_M}
$
as the speed of sound in the BEC regime. This reproduces
$
c_0 = \sqrt{U_M n_M/m_M}
$
in the dilute BEC limit where $n_M \to n/2$~\cite{Fetter, leggett}. 
Thus, for a proper description of the BEC physics, this analysis shows that 
it is sufficient to retain coefficients $A$, $B$, $Q_{ij}$, and $C_{ij}$ in 
the expansion of $\det \mathbf{M}_q$.

In the presence of a superflow when $\mathbf{Q \ne 0}$, the substitution
Eq.~(\ref{eqn:iql}) is equivalent to replacing $\omega$ with
$
\bar{\omega} = \omega - \sum_i s_\mathbf{k q}^i Q_i
$
in $\det \mathbf{M}_{q}$. Thus, in the BEC regime, we obtain
$
\det \mathbf{\bar{M}}_{\bar{q}, q_\ell=0} \approx 
\bar{A} \sum_{ij} \bar{C}_{ij} q_i q_j 
- \sum_{ij} \bar{B}^i_{\mathbf{0}^+}
\bar{B}^j_{\mathbf{0}^+} Q_i Q_j
+ (\sum_{ij} \bar{C}_{ij} q_i q_j)^2,
$
where $\bar{A}$ and $\bar{C}_{ij}$ are obtained by replacing
$\xi_\mathbf{k}$ with $\bar{\xi}_\mathbf{k}$ and
$E_\mathbf{k}$ with $\bar{E}_\mathbf{k}$, and
$
\bar{B}^i_\mathbf{q} = \frac{1}{N} \sum_\mathbf{k} 
\frac{\bar{\xi}_\mathbf{k}}{4\bar{E}_\mathbf{k}^3} s_\mathbf{k q}^i.
$
The origin of this coefficient can be traced back to 
$\bar{m}_{1, \mathbf{q}}^i$, which is given in Appendix~\ref{sec:app}.
Analogous to the analysis of $n_\mathrm{B}^d$ in Sec.~\ref{sec:bbc}, 
we approximate
$
s_\mathbf{k q}^i \approx \sum_j \frac{\partial^2 \xi_\mathbf{k}}
{\partial k_j \partial k_i} q_j
$
in the $\mathbf{q \to 0}$ limit, and approximate
$
\bar{W}_{\bar{q}, q_\ell = 0}
$
by its $\mathbf{q = 0}$ value, $\bar{W}_0$, leading to
$
\mathbf{\bar{W}}^\mathrm{T}_{\bar{q}}
\mathbf{\bar{M}}^{-1}_{\bar{q}} \mathbf{\bar{W}}_{\bar{q}}|_{q_\ell = 0} 
\approx 
\frac{2 \bar{W}_0^2 \sum_{ij} \bar{Q}_{ij} q_i q_j}
{\det \mathbf{\bar{M}}_{\bar{q}} |_{q_\ell=0}}.
$
Substituting this expression back into Eq.~(\ref{eqn:bOBd}), we find
\begin{align}
\label{eqn:DBdBEC}
D_{\mathrm{B}, d}^{ij} = \frac{\kappa}{N} \sum_\mathbf{q} 
\frac{2 W^2_0 B^i_{\mathbf{0}^+} 
B^j_{\mathbf{0}^+}}
{\sum_{ij} Q_{ij} q_i q_j (A+\sum_{ij} Q_{ij} q_i q_j)^2},
\end{align}
where
$
A = 4\Delta^2/U^3,
$
$
W_0 = 2\Delta/U^2,
$
and
$
Q_{ij} = (m_M^{-1})_{ij}/(2U^2)
$
in the dilute BEC limit. In addition, using integration by parts,
$
B^i_{\mathbf{0}^+} = - \frac{1}{N} \sum_{j \mathbf{k}} 
\frac{\partial}{\partial k_j} \Big(\frac{\xi_\mathbf{k}}
{4E_\mathbf{k}^3}\Big) \frac{\partial \xi_\mathbf{k}}
{\partial k_i} q_j,
$
we find
$
B^i_{\mathbf{0}^+} = \frac{2}{U^2} \sum_j (m_M^{-1})_{ij} q_j
$
in the dilute BEC limit. Alternatively, Eq.~(\ref{eqn:DBdBEC}) 
can be obtained directly from the second term of Eq.~(\ref{eqn:DBd}) 
in this limit. For example, in cubic lattices with
$
\varepsilon_\mathbf{k} = -2t \sum_i^{\dim} \cos(k_i a),
$
where $t$ is the hopping parameter and $\dim = (2,3)$ denotes 
the spatial dimension of the system, we obtain
$
s_\mathbf{k q}^i \approx 2a^2 t \cos(k_i a)\, q_i,
$
and a direct calculation yields
$
B^i_{\mathbf{0}^+} = 8a^2 t^2 q_i / U^3,
$
which is consistent with the expected result
$
(m_M^{-1})_{ij} = 2 a^2 t_M \delta_{ij},
$
where $t_M = 2t^2/U$ is the effective hopping of Cooper 
molecules~\cite{nsr85} .
Similar to the analysis of Eq.~(\ref{eqn:nBd}), in the dilute limit 
we can drop the $\mathbf{q}$-dependent term in the denominator, i.e., 
from the $(A + \sum_{ij} Q_{ij} q_i q_j)^2$ factor for the lattice model, 
and replace $q_i q_j$ with $|\mathbf{q}|^2/\dim$ in the numerator, 
leading to
$
D_{\mathrm{B}, d}^{ij} = \frac{8 n_\mathrm{B}^d}{\dim} (m_M^{-1})_{ij}.
$ 
Thus, we confirm that the superfluid depletion $D_{\mathrm{B}, d}^{ij}$ 
agrees with the known results in the dilute limit. For instance, 
in the continuum model where $m_M = 2m$, we reproduce the well-known result
$
m D_{\mathrm{B}, d} = 4 n_\mathrm{B}^d / \dim
$
~\cite{huang92, giorgini94, lopatin02, cappellaro19, orso07}. 
In comparison to the expression for atomic Bose superfluids, 
we note the presence of a factor of four when the above result is 
written in terms of the effective molecular mass and filling. 
This factor is analogous to the prefactor appearing in 
$D_{0,0} = 4 n_M / m_M$ in the BEC regime.
This demonstrates that random disorder depletes superfluid particles 
more strongly than it depletes condensate particles in the dilute BEC limit.

\subsection{Speed of Sound}
\label{sec:ss}

As discussed in Sec.~\ref{sec:dbl}, in the absence of disorder when 
$\kappa = 0$, the collective-mode spectrum can be obtained from the 
poles of the fluctuation propagator by setting $\det \mathbf{M}_q = 0$
~\cite{engelbrecht97, taylor06, diener08}. 
In the limit $\mathbf{q \to 0}$, these excitations correspond to 
phase oscillations of the order parameter and represent the Goldstone 
mode associated with the spontaneous breaking of the $U(1)$ symmetry 
in the superfluid state. In the $\mathbf{q \to 0}$ limit, this yields
$
\omega_\mathbf{q}^2 = \sum_{ij} v_0^{ij} q_i q_j,
$
characteristic of a sound mode, where the square root of 
\begin{align}
v_0^{ij} = \frac{A Q_{ij}}{AR + B^2}
\end{align}
determines its speed throughout the BCS-BEC crossover, e.g., 
$
v_0^{ij} = c_0^2 \delta_{ij}
$ 
for isotropic systems. In particular to the BEC regime, one finds
$
v_0^{ij} \approx A Q_{ij}/B^2 = U_M n_{c,0} (m_M^{-1})_{ij},
$
where $n_{c,0}$ is the condensate fraction~\cite{engelbrecht97, taylor06, diener08}.
If this expression is applied directly to the disordered case, it would 
suggest that the sound speed decreases, since $n_c$ decreases in comparison 
to $n_{c,0}$ with increasing $\kappa$ from 0. 
However, this is known to be incorrect for weakly-interacting atomic bosons. 
In other words, the usual relation does not hold in the presence of 
disorder, i.e., 
$
v_s^{ij} \ne U_M n_c (m_M^{-1})_{ij}
$
when $\kappa \ne 0$.

Given the successful reproduction of the thermodynamic properties in
the BEC regime, we follow the alternative recipe given by Ref.~\cite{giorgini94} 
and use the thermodynamic definition of the sound speed in superfluids,
\begin{align}
v_s^{ij} = D_{s}^{ij} \mathcal{X}_s^{-1},
\end{align}
where $D_{s}^{ij}$ is the superfluid fraction and 
$
\mathcal{X}_s = \frac{\partial n}{\partial \mu}\big|_{N}
$
is the static compressibility. Using 
$
\mu = \frac{\partial \mathcal{E}}{\partial n}\big|_N,
$
we may equivalently write
$
\mathcal{X}_s^{-1} = \frac{\partial^2 \mathcal{E}}{\partial n^2}\big|_N,
$
where
$
\mathcal{E} = \Omega + \mu n
$
is the ground-state energy per site at $T=0$.
Similar to our treatment of the disorder correction to the superfluid 
fraction,
$
D_{s}^{ij} = D_{s,0}^{ij} - D_d^{ij},
$
where $D_d^{ij}$ denotes the superfluid depletion due to disorder, 
we decompose the compressibility as
$
\mathcal{X}_s^{-1} = \mathcal{X}_0^{-1} + \mathcal{X}_d^{-1},
$
where $\mathcal{X}_0^{-1}$ is the $\kappa=0$ contribution and 
$\mathcal{X}_d^{-1}$ is the disorder correction. Our second-order 
perturbative expansion of the disorder potential remains controlled as 
long as $D_{s,0}^{ij} \gg D_d^{ij}$ and 
$\mathcal{X}_0^{-1} \gg \mathcal{X}_d^{-1}$. Within this construction, 
we may approximate
\begin{align}
\label{eqn:vsij}
v_s^{ij} = v_0^{ij} \bigg( 
1 - \frac{D_{d}^{ij}}{D_{0}^{ij}}
+ \frac{\mathcal{X}_d^{-1}}{\mathcal{X}_0^{-1}}
\bigg),
\end{align}
throughout the BCS-BEC crossover, where 
$
v_0^{ij} = D_{s,0}^{ij} \mathcal{X}_0^{-1}
$
coincides with the $\kappa = 0$ result discussed above.

To illustrate this in the dilute BEC limit, note that the disorder 
correction to the energy can be written as
$
\delta \mathcal{E} = \delta \Omega + n \delta \mu
$
at $T=0$. Noting that $\delta \mu \simeq 0$ in this limit, as the 
saddle-point condition Eq.~(\ref{eqn:spc}) does not depend explicitly 
on $\kappa$, we approximate
$
\mathcal{X}_d^{-1} \equiv \frac{\partial^2 \Omega_\mathrm{B}^d}{\partial n^2}\big|_N.
$
Proceeding as in Secs.~\ref{sec:bbc} and~\ref{sec:dbl}, we expand 
$W_{\mathbf{q},0}$ and 
$M^{11}_{\mathbf{q},0}+M^{12}_{\mathbf{q},0}$ in the $\mathbf{q\to 0}$ 
regime, which yields
$
\delta \mathcal{E} = -\frac{\kappa_M}{N} \sum_\mathbf{q} 
\frac{n_M}{2\mu_M + \frac{1}{2}\sum_{ij} (m_M^{-1})_{ij} q_i q_j}.
$
This is in agreement with the disorder correction for weakly-interacting 
bosons~\cite{huang92, giorgini94, lopatin02, cappellaro19}. 
As a further illustration, we consider isotropic systems, where
$
D_d^{ij} = D_d \delta_{ij}
$
and
$
D_0^{ij} = D_0 \delta_{ij},
$
and calculate
$
\mathcal{X}_d^{-1} = \frac{\partial^2 \delta \mathcal{E}}{\partial n^2}\big|_N,
$
leading eventually to
$
\frac{\mathcal{X}_d^{-1}}{\mathcal{X}_0^{-1}} = \frac{\dim^2}{4} 
\frac{D_d}{D_0}.
$
Thus, the sound speed is given by
$
c_s^2 - c_0^2 = \frac{\dim^2 - 4}{4} \frac{D_d}{D_0},
$
which reproduces the known continuum result for weakly-interacting 
atomic bosons in three dimensions~\cite{giorgini94}, where, showing that the 
sound speed increases with disorder. 
Here, we used
$
\sum_\mathbf{q} |\mathbf{q}|^2/\theta_\mathbf{q}^6 = 
\frac{\dim}{4}  \sum_\mathbf{q} 1/\theta_\mathbf{q}^4,
$
where
$
\theta_\mathbf{q} = \sqrt{|\mathbf{q}|^2 + 4m_M \mu_M}.
$
Eq.~(\ref{eqn:vsij}) shows that this 
is caused by the corresponding increase in the inverse static compressibility, 
which overcompensates the depletion of the superfluid fraction.

\section{Numerical Illustration}
\label{sec:ni}

\begin{figure} [htb]
\centerline{\scalebox{0.63}{\includegraphics{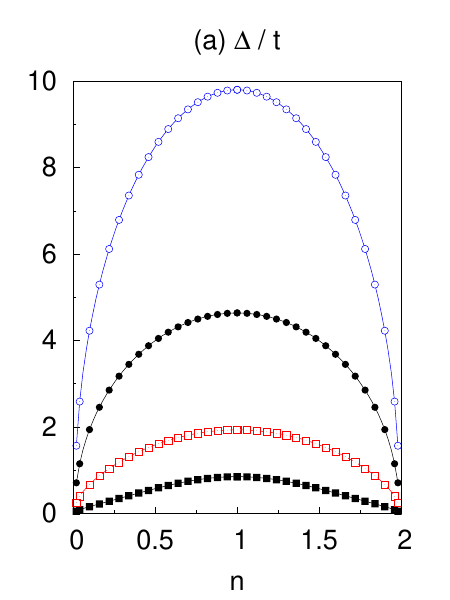} 
\hskip -0.8cm
\includegraphics{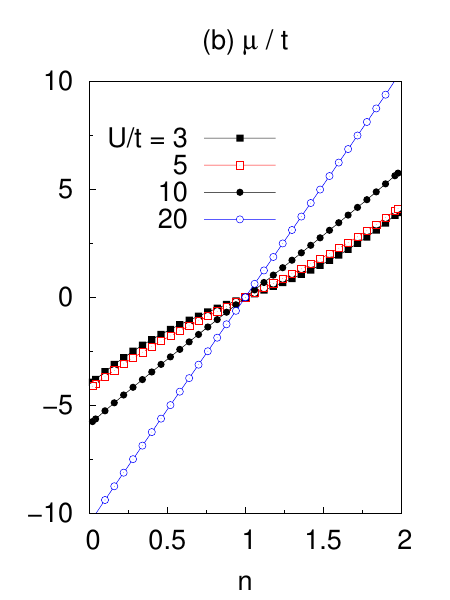}}}
\caption{\label{fig:gap} 
Self-consistent solutions for (a) the order parameter $\Delta$ 
and (b) the chemical potential $\mu$ as functions of the particle 
filling $n$ for $\kappa = 0.1t^2$. 
The results exhibit exact particle-hole symmetry about half filling 
($n = 1$); therefore, only the particle side of the filling range 
is shown in Fig.~\ref{fig:fig2}.
}
\end{figure}

\begin{figure*} [htb]
\includegraphics[width = 0.99\textwidth]{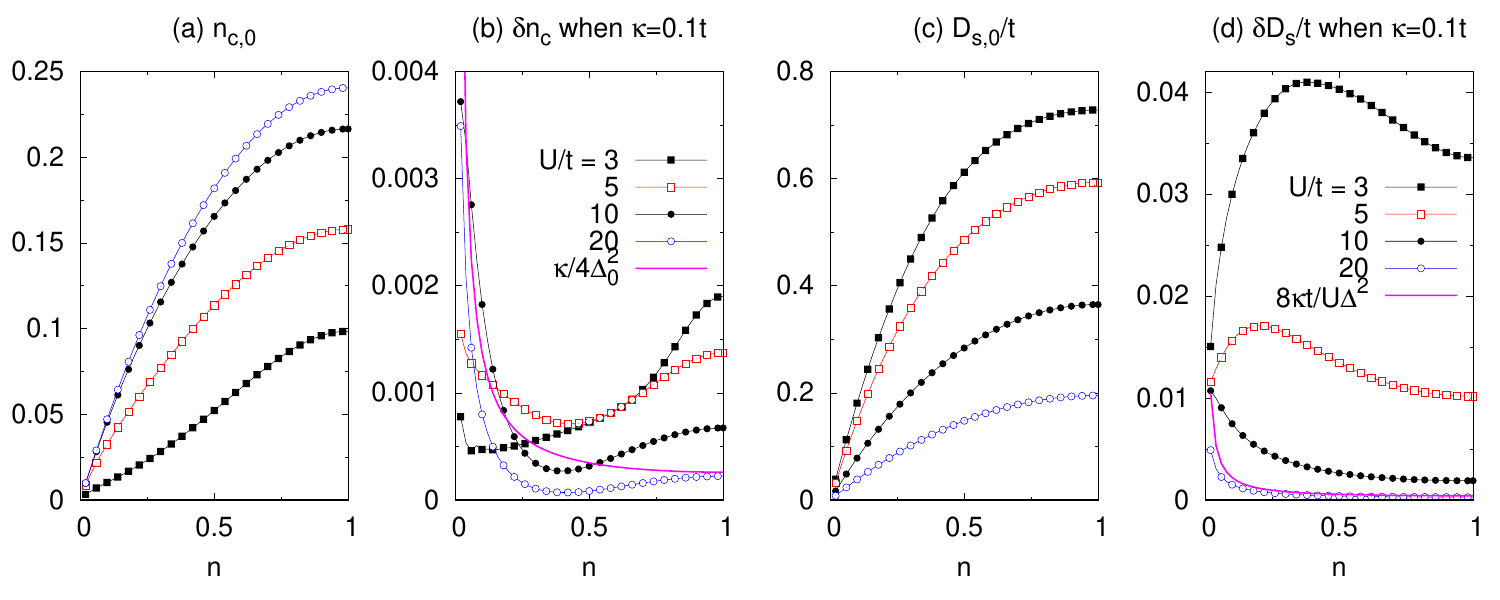}
\caption{\label{fig:fig2} 
In (a) the condensate fraction $n_{c,0}$, and in (c) the superfluid 
fraction $D_{s,0}$ are shown as functions of the filling 
$0 \lesssim n \leq 1$ in the absence of disorder ($\kappa = 0$). 
For comparison, panels (b) and (d) display the corresponding 
disorder-induced reductions, $\delta n_c = n_{c,0} - n_c$ and 
$\delta D_s = D_{s,0} - D_s$, respectively, for weak disorder 
($\kappa = 0.1t^2$). 
}
\end{figure*}

As an illustration, we consider the square lattice with dispersion
\begin{align}
\varepsilon_\mathbf{k} = -2t \cos(k_x a) - 2t \cos(k_y a),
\end{align}
where $t \geq 0$ denotes the nearest-neighbor hopping amplitude. 
The momentum components $k_x$ and $k_y$ lie within the Brillouin zone, 
$-\pi/a \leq k_x < \pi/a$ and $-\pi/a \leq k_y < \pi/a$. 
We first employ an iterative approach to solve 
Eqs.~(\ref{eqn:spc}) and~(\ref{eqn:n}) self-consistently for 
$\Delta$ and $\mu$ at given parameters 
$0 \leq n \leq 2$, $U/t \geq 0$ and $\kappa/t \geq 0$. 
For example, numerical results for $\Delta$ and $\mu$ are shown 
in Fig.~\ref{fig:gap} as functions of $n$ for $U/t = \{3,5,10,20\}$. 
Due to convergence issues in the iterative scheme, the numerical 
accuracy becomes unreliable in the dilute ($n \to 0$) and 
band-insulator ($n \to 2$) limits where $\Delta/t \ll 1$, 
and these regimes are therefore not displayed. The results exhibit 
exact particle-hole symmetry about half filling ($n = 1$). 
The self-consistent solutions are then substituted into 
Eqs.~(\ref{eqn:nc0}) and~(\ref{eqn:ncd}), together with 
Eqs.~(\ref{eqn:D0}),~(\ref{eqn:DFd}) and~(\ref{eqn:DBd}), 
to determine the condensate and superfluid fractions. 
The numerical results are summarized in Fig.~\ref{fig:fig2}, where 
the superfluid-weight tensor reduces to 
$
D_s^{ij} = D_s \delta_{ij},
$
reflecting isotropic diagonal elements enforced by $C_4$ symmetry. 
Moreover, owing to particle-hole symmetry about $n = 1$, there 
we present results only for $0 \lesssim n \leq 1$.


We emphasize that our perturbative results are valid only in the 
weak-disorder regime. Depending on $n$, the theory eventually breaks down 
once $\kappa/t$ becomes sufficiently large for a given $U/t$. Likewise, 
the theory also fails when $U/t$ becomes too small toward the BCS regime 
at fixed $\kappa/t$. Therefore, for a given $U/t$, the approach is expected 
to be more reliable at larger $\kappa/t$ values away from the dilute limit, 
where $\Delta/t$ is enhanced. For the number equation, our numerical results 
show that, for fixed $U/t$ and $n$, increasing $\kappa/t$ gradually reduces 
$n_\mathrm{F}^0$ and consequently increases $n_\mathrm{F}^d + n_\mathrm{B}^d$ 
in the dilute limit, while having little effect near half filling. 
In addition, whereas $n_\mathrm{F}^d$ typically dominates over $n_\mathrm{B}^d$ 
in the BCS regime, the opposite holds in the BEC regime. 
Outside the dilute limit, this balance causes $\Delta$ to remain 
essentially unaffected by weak disorder, even in the BEC regime, 
particularly near half filling.

In Figs.~\ref{fig:fig2}(a) and~\ref{fig:fig2}(b), we show the disorder-free 
condensate fraction $n_{c,0}$ and its disorder-induced correction 
$\delta n_c = n_{c,0} - n_c$ for $\kappa = 0.1 t^2$. 
Our numerical results are in excellent agreement with the expectation that 
$n_{c,0}$ is exponentially suppressed in the BCS regime and saturates to 
$n(2-n)/4$ in the BEC regime, which approaches $n/2$ in the dilute limit.
For $\kappa \ne 0$, we also verified that
$
\delta n_c \to (\Delta_0^2 - \Delta^2)/U^2,
$
consistent with our analytical expectations in the dilute BEC limit, 
where $n_{c,0} - n_c^0 \gg n_c^d$. 
For instance, the analytical result $\kappa / (4 \Delta_0^2)$ is shown 
in Fig.~\ref{fig:fig2}(b) for $U/t = 20$, which is in good agreement with 
the numerical results not only in the dilute limit but also around 
half filling. We also confirm that the agreement becomes much better as 
$U/t$ increases. In addition, we find that $\delta n_c$ generally scales 
with $\kappa / \Delta_0^2$ in the BEC regime.
In Figs.~\ref{fig:fig2}(c) and~\ref{fig:fig2}(d), we show the disorder-free 
superfluid fraction $D_{s,0}$ and its disorder-induced correction 
$\delta D_s = D_{s,0} - D_s$ for $\kappa = 0.1 t^2$. 
Our numerical results are consistent with the expectations that
$
D_{s,0} = n/m_*
$
in the dilute BCS limit, where $m_* = 1/(2ta^2)$ is the effective band 
mass of the particles, and
$
D_{s,0} = 4 n_{c,0} / m_M
$
in the BEC regime, where $m_M = U/(4t^2 a^2)$ is the effective band mass 
of the Cooper molecules. Note that
$
D_{s,0} = 4 n_M/m_M
$
in the dilute BEC limit, where $n_M = n/2$ is the filling of Cooper molecules.
For $\kappa \ne 0$, we further verified that 
$
\delta D_s \to 8\kappa t^2/(U \Delta^2)
$ 
not only in the dilute BEC limit. 
For instance, this result is shown in Fig.~\ref{fig:fig2}(d) for $U/t = 20$, 
which is in good agreement with the numerical results not only in the 
dilute limit but also around half filling.
Away from the dilute limit, Figs.~\ref{fig:fig2}(b) and~\ref{fig:fig2}(d) 
show that the disorder-induced corrections to both the condensate and 
the superfluid fractions become small in the BEC regime. Thus, superfluidity 
is more robust against disorder near half filling in the BEC regime.

Although not shown in the figures, we note that when the theory is pushed 
beyond the weak-disorder regime by increasing $\kappa/t$ at fixed $U/t$, 
the superfluid fraction $D_s$ eventually becomes negative, first in the dilute 
limit and subsequently at higher fillings. Moreover, larger $U/t$ values 
require proportionally larger $\kappa/t$ for $D_s$ to change sign. 
For instance, for $n = 0.875$ of Ref.~\cite{ghosal01}, $D_s$ changes sign 
at $\kappa/t \approx \{0.75, 2.05, 3.70\}$ when $U/t = \{2,3,4\}$. 
According to Eq.~(\ref{eqn:Oeff}), $D_s$ measures the response of the 
superfluid system to a phase twist of the order parameter, i.e., the 
change in the thermodynamic potential under an infinitesimal superfluid flow. 
Thus, the stability of a spatially uniform superfluid requires a 
positive-definite superfluid-fraction tensor, since a negative eigenvalue 
indicates that the superconducting state is unstable toward the spontaneous 
formation of a phase gradient, i.e., a spatially nonuniform superfluid. 
This interpretation is consistent with Bogoliubov-de Gennes calculations, 
which demonstrate that the pairing amplitude becomes increasingly 
inhomogeneous with stronger disorder~\cite{ghosal01}. In the strong-disorder 
regime, the system was found to fragment into superconducting islands 
characterized by a large pairing amplitude, embedded within an insulating 
background.

While the present analysis captures pairing-dominated physics in the 
low- and intermediate-filling regimes, we note that near half-filling, 
perfect nesting of the Fermi surface can promote competing 
charge-density-wave order. Quantum Monte Carlo studies of the 
half-filled attractive Hubbard model have shown that superconductivity 
and charge ordering coexist in the clean limit, but that disorder 
immediately destroys the charge order while leaving pairing correlations robust~\cite{huscroft97}. Therefore, the Gaussian fluctuation approach 
employed here is not intended to describe that regime, where density 
fluctuations become significant. To the best of our knowledge, the 
interplay between pairing and density-wave channels has not been 
systematically explored even in the clean BCS-BEC crossover, making 
it an interesting avenue for future study.

\section{Conclusion}
\label{sec:conc}

In this work, we developed a systematic and controlled theoretical approach 
to study the effects of weak quenched disorder on the BCS-BEC crossover within 
the Hubbard model. Using a functional-integral formalism, we derived the 
thermodynamic potential up to second order in both the disorder potential 
and pairing fluctuations, allowing us to obtain self-consistent analytic 
expressions for key zero-temperature properties, including the number equation, 
condensate fraction, superfluid fraction and sound speed. 
A crucial validation of our approach lies in its ability to analytically 
recover the well-known continuum limits of weakly-interacting bosons in the 
dilute BEC limit~\cite{huang92, giorgini94, lopatin02, cappellaro19, orso07}. 
In this limit, we find that weak disorder, by breaking translational symmetry, 
depletes the superfluid fraction more strongly than the condensate 
fraction and increases the sound velocity due to an overcompensation of the 
static compressibility. These results provide a unified and robust theoretical 
approach for describing the BCS-BEC crossover in disordered lattice systems.

Looking ahead, our formalism provides a solid basis for several important 
extensions. In particular, extending this approach to finite temperatures 
would enable direct comparison with experiments on ultracold atomic gases 
and provide new insights into the behavior of disordered
superconductors~\cite{iskin25e}. 
The present results capture the leading-order disorder corrections 
valid in the weak-disorder limit, while the regime of stronger disorder, 
where localization or non-Gaussian effects become relevant, remains an 
interesting direction for future work. 
Moreover, the approach can be naturally generalized to more complex systems, 
such as multiband Hubbard models, offering a powerful route to investigate 
the interplay between disorder and quantum-geometric effects, including the 
superfluid weight, in strongly-correlated 
materials~\cite{torma22, yu24, jiang25, lau22, chan25, bouzerar25}. 
The latter direction will be the focus of our planned follow-up work.

\begin{acknowledgments}
We thank G. Orso for valuable email correspondence and insightful 
input on several aspects of our discussions. The author acknowledges 
funding from US Air Force Office of Scientific Research (AFOSR) 
Grant No. FA8655-24-1-7391.
\end{acknowledgments}

\appendix

\section{Derivation of $D_{\mathrm{B}, d}^{ij}$ given in Eq.~(\ref{eqn:DBd})}
\label{sec:app}

Since 
$
\bar{\Omega}_\mathrm{B}^d \equiv \bar{\Omega}_\mathrm{B}^d(\bar{\mu}, \mathbf{Q})
$
and 
$
D_{\mathrm{B}, d}^{ij} =
- \frac{\partial^2 \bar{\Omega}_\mathrm{B}^d}
{\partial Q_i \partial Q_j}\big|_{\bar{\mu}, \mathbf{Q = 0}},
$
we only need to track the $\mathbf{Q}$ dependence arising from the
$\bar{q}_\ell$ terms when $q_\ell = 0$. For this purpose, it is convenient 
to expand the matrix element as
$
\bar{M}^{11}_{\bar{q}, q_\ell = 0} = \bar{m}_{0, \mathbf{q}} 
+ \sum_i\bar{m}_{1, \mathbf{q}}^i Q_i
+ \sum_{ij} \bar{m}_{2,\mathbf{q}}^{ij} Q_i Q_j + \cdots,
$
up to second order in $\mathbf{Q}$, where
\begin{align}
\bar{m}_{0, \mathbf{q}} &= \frac{1}{U} - \frac{1}{N} \sum_\mathbf{k} 
\frac{\bar{E}\bar{E}' + \bar{\xi}\bar{\xi}'}
{2\bar{E}\bar{E}'(\bar{E}+\bar{E}')}, \\
\bar{m}_{1, \mathbf{q}}^i &= \frac{1}{N} \sum_\mathbf{k} 
\frac{\bar{E}\bar{\xi}' + \bar{E}'\bar{\xi}}
{2\bar{E}\bar{E}'(\bar{E}+\bar{E}')^2} s_\mathbf{k q}^i, \\
\bar{m}_{2,\mathbf{q}}^{ij} &= -\frac{1}{N} \sum_\mathbf{k} 
\frac{\bar{E}\bar{E}' + \bar{\xi}\bar{\xi}'}
{2\bar{E}\bar{E}'(\bar{E}+\bar{E}')^3} s_\mathbf{k q}^i s_\mathbf{k q}^j.
\end{align}    
Similarly, we expand
$
\bar{M}^{12}_{\bar{q}, q_\ell = 0} = \bar{o}_{0, \mathbf{q}} 
+ \sum_{ij} \bar{o}_{2,\mathbf{q}}^{ij} Q_i Q_j + \cdots,
$
where
\begin{align}
\bar{o}_{0, \mathbf{q}} &= \frac{1}{N} \sum_\mathbf{k} 
\frac{\Delta^2}
{2\bar{E}\bar{E}'(\bar{E}+\bar{E}')}, \\
\bar{o}_{2,\mathbf{q}}^{ij} &= \frac{1}{N} \sum_\mathbf{k} 
\frac{\Delta^2}
{2\bar{E}\bar{E}'(\bar{E}+\bar{E}')^3} s_\mathbf{k q}^i s_\mathbf{k q}^j.
\end{align}    
The coupling term between bosonic fluctuations and the disorder 
potential is expanded as
$
\bar{W}_{\bar{q}, q_\ell = 0} = \bar{w}_{0, \mathbf{q}} 
+ \sum_i\bar{w}_{1, \mathbf{q}}^i Q_i
+ \sum_{ij} \bar{w}_{2,\mathbf{q}}^{ij} Q_i Q_j + \cdots,
$
where
\begin{align}
\bar{w}_{0, \mathbf{q}} &= - \frac{1}{N} \sum_\mathbf{k} 
\frac{\Delta(\bar{\xi}+\bar{\xi}')}
{2\bar{E}\bar{E}'(\bar{E}+\bar{E}')}, \\
\bar{w}_{1, \mathbf{q}}^i &= \frac{1}{N} \sum_\mathbf{k} 
\frac{\Delta}
{2\bar{E}\bar{E}'(\bar{E}+\bar{E}')} s_\mathbf{k q}^i, \\
\bar{w}_{2,\mathbf{q}}^{ij} &= -\frac{1}{N} \sum_\mathbf{k} 
\frac{\Delta(\bar{\xi}+\bar{\xi}')}
{2\bar{E}\bar{E}'(\bar{E}+\bar{E}')^3} s_\mathbf{k q}^i s_\mathbf{k q}^j.
\end{align}    
Using these expansions, we find that
$
\det \mathbf{\bar{M}}_{\bar{q}, q_\ell = 0} = \bar{d}_{0, \mathbf{q}} 
+ \sum_{ij} \bar{d}_{2, \mathbf{q}}^{ij} Q_i Q_j + \cdots
$
up to second order in $\mathbf{Q}$, where
$
\bar{d}_{0, \mathbf{q}} = \bar{m}_{0, \mathbf{q}}^2 - \bar{o}_{0, \mathbf{q}}^2
$
and
$
\bar{d}_{2, \mathbf{q}}^{ij} = 2\bar{m}_{0, \mathbf{q}} \bar{m}_{2,\mathbf{q}}^{ij}
- 2\bar{o}_{0, \mathbf{q}} \bar{o}_{2,\mathbf{q}}^{ij}
- \bar{m}_{1, \mathbf{q}}^i \bar{m}_{1, \mathbf{q}}^j.
$
Similarly, we obtain
$
\mathbf{\bar{W}}^\mathrm{T}_{\bar{q}}
\mathbf{\bar{M}}^{-1}_{\bar{q}} \mathbf{\bar{W}}_{\bar{q}}|_{q_\ell = 0} 
= \frac{\bar{e}_{0, \mathbf{q}}}{\bar{d}_{0, \mathbf{q}}} 
+ \sum_{ij} \Big(
\frac{\bar{e}_{2, \mathbf{q}}^{ij}} {\bar{d}_{0, \mathbf{q}}} 
- \frac{\bar{e}_{0, \mathbf{q}} \bar{d}_{2, \mathbf{q}}^{ij}}
{\bar{d}_{0, \mathbf{q}}^2} 
\Big) Q_i Q_j + \cdots
$
again up to second order in $\mathbf{Q}$, where
$
\bar{e}_{0, \mathbf{q}} = 2(\bar{m}_{0, \mathbf{q}}-\bar{o}_{0, \mathbf{q}})
\bar{w}_{0, \mathbf{q}}^2
$
and
$
\bar{e}_{2, \mathbf{q}}^{ij} = 2(\bar{m}_{0, \mathbf{q}}+\bar{o}_{0, \mathbf{q}})
\bar{w}_{1, \mathbf{q}}^i \bar{w}_{1, \mathbf{q}}^j
-4 \bar{w}_{0, \mathbf{q}} \bar{m}_{1, \mathbf{q}}^i \bar{w}_{1, \mathbf{q}}^j
+4 (\bar{m}_{0, \mathbf{q}}-\bar{o}_{0, \mathbf{q}}) \bar{w}_{0, \mathbf{q}}
\bar{w}_{2,\mathbf{q}}^{ij}
+2 \bar{w}_{0, \mathbf{q}}^2(\bar{m}_{2,\mathbf{q}}^{ij}-\bar{o}_{2,\mathbf{q}}^{ij}).
$
In the final expression given in Eq.~(\ref{eqn:DBd}), all coefficients are
evaluated in the absence of a superflow, i.e.,
$
d_{0, \mathbf{q}} = \bar{d}_{0, \mathbf{q}}|_\mathbf{Q = 0},
$
$
d_{2, \mathbf{q}}^{ij} = \bar{d}_{2, \mathbf{q}}^{ij}|_\mathbf{Q = 0},
$
$
e_{0, \mathbf{q}} = \bar{e}_{0, \mathbf{q}}|_\mathbf{Q = 0}
$
and
$
e_{2, \mathbf{q}}^{ij} = \bar{e}_{2, \mathbf{q}}^{ij}|_\mathbf{Q = 0}.
$

\bibliography{refs}

\end{document}